\documentclass[journal]{IEEEtran}
\ifCLASSINFOpdf
\else
\fi

\usepackage{graphicx}          
\usepackage{amsmath,amsfonts,amssymb}

\DeclareMathOperator{\sgn}{sgn}

\newtheorem{remark}{Remark}

\newtheorem{property}{Property}
\newtheorem{proof}{Proof}
\newtheorem{theorem}{Theorem}

\newtheorem{lemma}{Lemma}

\newtheorem{defn}{Definition}

\newcommand{\cfbox}[2]{%
	\colorlet{currentcolor}{.}%
	{\color{#1}%
		\fbox{\color{currentcolor}#2}}%
}

\usepackage{pbox}
\usepackage{breqn}
\usepackage{widetext}
\usepackage{stfloats}
\usepackage{xcolor}
\usepackage{graphicx}
\usepackage{cite}
\usepackage{picinpar}
\usepackage{amsmath}
\usepackage{stfloats}
\usepackage{url}
\usepackage{flushend}
\usepackage[latin1]{inputenc}
\usepackage{colortbl}
\usepackage{soul}
\usepackage{multirow}
\usepackage{pifont}
\usepackage{color}
\usepackage{alltt}
\usepackage[hidelinks]{hyperref}
\usepackage{enumerate}
\usepackage{siunitx}
\usepackage{breakurl}
\usepackage{epstopdf}
\usepackage{nomencl} 
\makenomenclature

\begin{document}
\title{On-line Capacity Estimation for Lithium-ion Battery Cells via an Electrochemical Model-based Adaptive Interconnected Observer}

\author{Anirudh~Allam,~\IEEEmembership{Student~Member,~IEEE,}
	Simona~Onori,~\IEEEmembership{Senior~Member,~IEEE}
	\thanks{A. Allam and S. Onori are with the Energy Resources Engineering department, Stanford University, Stanford, CA 94305, USA (e-mail: aallam@stanford.edu; sonori@stanford.edu). (\textit{Corresponding author: Simona Onori.})}
}
%
%

\markboth{}%
{Shell \MakeLowercase{\textit{et al.}}: Bare Demo of IEEEtran.cls for IEEE Journals}
%



\maketitle
\begin{abstract}
	Battery aging is a natural process that contributes to capacity and power fade, resulting in a gradual performance degradation over time and usage. State of Charge (SOC) and State of Health (SOH) monitoring of an aging battery poses a challenging task to the Battery Management System (BMS) due to the lack of direct measurements. Estimation algorithms based on an electrochemical model that take into account the impact of aging on physical battery parameters can provide accurate information on lithium concentration and cell capacity over a battery's usable lifespan. %
	\textcolor{black}{
	A temperature-dependent} electrochemical model, the Enhanced Single Particle Model (ESPM), forms the basis for the synthesis of an adaptive interconnected observer that exploits the relationship between capacity and power fade, due to the growth of Solid Electrolyte Interphase layer (SEI), to enable combined estimation of states (lithium concentration in both electrodes and cell capacity) and aging-sensitive transport parameters (anode diffusion coefficient and SEI layer ionic conductivity). The practical stability conditions for the adaptive observer are derived using Lyapunov's theory. Validation results against experimental data show a bounded 
	capacity estimation error 
	within 2$\%$ of 
	its true value. 
	Further, effectiveness of capacity estimation is tested for two cells at different stages of aging. Robustness of capacity estimates under measurement noise and sensor bias are studied.
\end{abstract}


\section*{Nomenclature}
\addcontentsline{toc}{section}{Nomenclature}
\begin{IEEEdescription}[\IEEEusemathlabelsep\IEEEsetlabelwidth{$V_1,V_2,V_3$}]
	\item[$A$] Cell cross sectional area [\text{m$^{2}$}].
	\item[$D_{e}$] Electrolyte phase diffusion [\text{m$^{2}$/s}].	
	\item[$D^{eff}_{e}$] Effective electrolyte phase diffusion [\text{m$^{2}$/s}].		
	\item[$D_{s,j}$] Solid phase diffusion [\text{m$^{2}$/s}].
	\item[$D_{s,j, ref}$] Reference solid phase diffusion [\text{m$^{2}$/s}].		
	\item[$E_{a}$] Activation energy.			
	\item[$F$] Faraday's constant [\text{C/mol}].
	\item[$L_{j}$] Domain thickness [\text{m}].
	\item[$L_{sei}$] SEI layer thickness [\text{m}].	
	\item[$M_{sei}$] Molar mass of SEI layer [\text{kg/mol}].	
	\item[$R_{g}$] Universal gas constant [\text{J/mol-K}].
	\item[$R_{j}$] Particle radius [\text{m}].
	\item[$R_{l}$] Lumped resistance [\text{$\Omega$}].
	\item[$R_{sei}$] SEI layer resistance [\text{$\Omega$}].	
	\item[$T_{ref}$] Reference temperature [\text{$^o$C}].	
	\item[$U_{j}$] Open circuit potential [\text{V}].	
	\item[$a_{s,j}$] Specific interfacial surface area [\text{m$^{-1}$}].	
	\item[$c_{e}$] Electrolyte phase concentration [\text{mol/m$^3$}].
	\item[$c_{e,0}$] Average electrolyte phase concentration [\text{mol/m$^3$}].
	\item[$c_{s,j}$] Solid phase concentration [\text{mol/m$^3$}].
	\item[$c_{s,j,bulk}$] Bulk concentration [\text{mol/m$^3$}].
	\item[$c_{s,j,surf}$] Surface concentration [\text{mol/m$^3$}].
	\item[$c_{s,j,max}$] Maximum solid phase concentration [\text{mol/m$^3$}].
	\item[$i_{s}$] Side reaction current density [\text{A/m$^2$}].
	\item[$k_{j}$] Reaction rate constant [\text{m$^{2.5}$/s-mol$^{0.5}$}].
	\item[$l$] Cartesian coordinate along the cell's thickness	
	\item[$r$] Radial coordinate.	
	\item[$t^{+}_{0}$] Transference number.	
	\item[$\phi_{e}$] Electrolyte potential [\text{V}].
	\item[$\epsilon_j$] Active volume fraction of solid phase.	
	\item[$\epsilon_{e,j}$] Porosity.
	\item[$\epsilon_{j,f}$] Active volume fraction of filler/binder.	
	\item[$\kappa$] Electrolyte conductivity [\text{S/m}].
	\item[$\kappa_{sei}$] SEI layer ionic conductivity [\text{S/m}].	
	\item[$\kappa^{eff}$] Effective electrolyte conductivity [\text{S/m}].	
	\item[$\eta_j$] Overpotential [\text{V}].
	\item[$\theta_{j,100\%}$] Reference stoichiometry ratio at 100$\%$ SOC.
	\item[$\theta_{j,0\%}$] Reference stoichiometry ratio at 0$\%$ SOC.
	\item[$\rho_{sei}$] SEI layer density.			
	\item[Subscript $j$] refers to anode, separator, or cathode.
	\item[Subscript $ol$] refers to open loop.
\end{IEEEdescription}

%
\IEEEpeerreviewmaketitle

\section{Introduction}
\IEEEPARstart{A}{ging} contributes to the diminishing performance in 
batteries, resulting in reliability and safety issues. %
It manifests in the form of energy and power fade, characterized by loss in cell capacity and increased internal impedance, respectively.
With respect to automotive applications, energy and power fade 
of a lithium-ion battery 
relates to a reduced driving range and limited acceleration performance at the vehicle level. 
The outcome of the work presented in this paper contributes towards accurate electrochemical model-based estimation of lithium concentration and cell capacity that holds the potential to enable the practical realization of advanced battery health-based control algorithms  in the future. 

\subsection{Background and Related Work}
There are various complex chemical and physical aging mechanisms affecting the anode and cathode in a battery \cite{vetter2005ageing}.  It is neither feasible to mathematically model the underlying dynamics of every aging mechanism, or their respective nonlinear interaction, with current technology and understanding, nor is it computationally viable to implement them in a control-oriented fashion. Most of the literature focuses on the Solid Electrolyte Interphase (SEI) layer growth, and considers it to be the dominant aging mechanism in lithium-ion batteries \cite{pinson2013theory}, \cite{prada2012simplified}. The SEI layer is a thin film formed around the active material in the negative electrode due to electrolyte decomposition that consumes cyclable lithium ions. With usage, the SEI layer grows gradually, not only causing capacity fade, but also resulting in power fade due to the increasing thickness of the layer, and the adverse effects of modified porosity on the effective transport properties in the electrolyte phase \cite{prada2012simplified}.%

Model-based  estimation of battery SOC and SOH is a well researched topic. Since aging affects the physical battery parameters, it is important to be aware that utilizing a fixed-parameter model for estimation purposes will yield estimates that will slowly diverge over time and usage. 
One way to counter this is to use dynamic aging models to keep track of the battery SOH. Physics-based aging models are not a viable option for real-time implementation due to the complexity that results from the lack of comprehensive knowledge of the various electrochemical aging mechanisms and their slow time-scale behavior. Whereas, semi-empirical aging models in combination with a battery Equivalent Circuit Model offers lower complexity at the cost of accuracy. However, such models also require extensive data for calibration and the accuracy of the model is not guaranteed as battery ages, unless the order of the model is increased significantly \cite{ahmed2015model}.  This has been the motivation to develop adaptive observers that update the parameters dynamically with aging.
Adaptive observers based on equivalent circuit models \cite{chaoui2015lyapunov}, \cite{du2016adaptive} operate by adapting the circuit parameters (resistors, capacitors) as aging progresses. 
On the other hand, an electrochemical model, such as the Single Particle Model (SPM), captures the concentration states and its parameters represent actual physical  properties. The SPM is a reduced-order electrochemical model that approximates each electrode by a spherical particle and neglects lithium concentration and migration dynamics in electrolyte phase, making it suitable for control-oriented applications. Electrochemical model-based adaptive estimation has provided promising results \cite{moura2014adaptive},  and \cite{jenkins2019fast}. In these algorithms, the lithium concentration states are estimated along with aging-sensitive parameters such as cyclable lithium ions, diffusion coefficient, and internal resistance. However, there has been no attempt at relating the results of the parameter estimates to the actual cell capacity. This is because the estimation algorithms do not incorporate the aging mechanisms into the modeling framework or relate the aging mechanisms to the changes observed in aging-sensitive parameters. Further, these adaptive estimation algorithms validate their functionality over a fresh cell and do not present any results against experimental datasets for an aged cell. In summary, the main contributions of this paper are motivated from the following shortcomings in the literature: 1) Semi-empirical aging models require large experimental datasets and are limited to the operating conditions they have been characterized for;
2) Model-based adaptive estimation algorithms, both equivalent circuit and electrochemical model-based, do not relate the estimated aging-sensitive parameters to the exact cell capacity or state of health. Clearly, there is a need for a framework that unites the strengths of adaptive estimation theory with physics-based modeling insights of degradation mechanisms, without relying on extensive experimental data for aging characterization or causing additional computational burden, and yet be able to predict battery capacity (SOH) in real time. Further, the framework must be general enough to be extended to varied battery chemistry and also allow other degradation mechanisms to be incorporated, if need be. 
\subsection{Contributions and Paper Outline}
The main contribution of this paper lies in exploiting the physico-chemical effects of the SEI layer growth on capacity and power fade, and combining it with the adaptive estimation theory in order to estimate the total cell capacity, lithium concentration, and aging-sensitive parameters in real time.
An aging-dependent voltage loss term that reflects the SEI layer-induced degradation is incorporated to enable the model to be used as the cell ages.  Further, a practically stable adaptive observer is implemented in a novel interconnected sliding mode observer structure in the presence of bounded modeling uncertainties, and validated against experimental data. Taking the practical limitations into considerations, such as inherent bounded uncertainties in the model, observability issues, and moderate sensitivities of the parameters to measured output variables, the trajectories of state and parameter estimates may not converge asymptotically to the true values. Hence, the notion of practical stability is explored for the proposed adaptive observer.
The remainder of this article is organized as follows: Section \ref{sec:Notations} details the notations and definitions used in the paper. Section \ref{sec:Echem_model} describes the ESPM and derives the coupling between capacity and power fade due to SEI layer growth. The state space representation of the ESPM (with aging induced effects) is formulated. The relationship between capacity and power fade motivates the design of an SPM-based adaptive interconnected sliding mode observer for the estimation of i) lithium concentration in electrodes, ii) aging-sensitive parameters, and iii) cell capacity in Section \ref{sec:Adaptive_Obs}. The practical stability of the estimation error dynamics is rigorously proved using Lyapunov's theory. Section \ref{sec:Results_Discussion} validates the proposed SPM-based adaptive observer against experimentally measured data, and Section \ref{sec:Conclusion} summarizes the conclusions.
\section{Preliminaries}\label{sec:Notations}
The following notations and symbols are used in the paper:
\begin{itemize}
	\item $||{\cdot}||$ is the Euclidean norm;
	\item $\mathbb{R}_{+} = \{z \in \mathbb{R} : z > 0 \}$; $\mathbb{R}_{-} = \{z \in \mathbb{R} : z < 0 \}$;	
	\item Matrix $C \in \mathbb{R}^{1\times n} $ is the output distribution vector defined as $C = [0 \hspace{0.1cm} 0 \dots 1]$;
	%
	%
	\item $\mathbb{B}_{||\epsilon||}$ is the ball of radius $||\epsilon||$ centered at the origin.	
	\item $I_{n}$ is the identity matrix of order $n$.
	\item Subscript $j$ denotes the domain in the lithium-ion battery. In the solid phase, it denotes the negative and positive electrode, $j \in [n,p]$. Whereas, in the electrolyte phase, it represents the negative electrode, separator, and positive electrode, $j\in[n,s,p]$.
\end{itemize}

\begin{defn} \label{defn:pers_excite} \cite{ioannou1996robust}
	A function $w:\mathbb{R}_{+} \mapsto \mathbb{R}^{n}$ is persistently exciting if there exist $T, \delta_{1}, \delta_{2} > 0$ such that
	\begin{equation}
		\delta_{1}I_{n} \le \int_{t}^{t+T} w(\tau)w^{T}(\tau) d\tau \le \delta_{2}I_{n} \nonumber
	\end{equation}
	holds for all $t \ge 0$.
\end{defn}
\begin{defn}\label{defn:Lipschitz_fun}
	A function $f(z,t) : \mathbb{R} \times \mathbb{R} \rightarrow \mathbb{R} $ is said to be globally Lipschitz in $z$ and uniformly in $t$ if, for some constant $L \in \mathbb{R}_{+}$, 
	$||f(z+\delta_{z}, t) - f(z, t)|| \le L ||\delta_{z}||$ holds true.
\end{defn}

\begin{defn} \label{defn:practical_stab}
	\cite{lakshmikantham1990practical} A dynamic system $\dot{z} = f\left(t,z\right)$ with initial condition $z(t_{0}) = z_{0}$ 
	is practically stable if $\exists$ $a, b$ with $0 < a < b$ such that $||z_{0}|| < a $ and $||z(t)|| < b$, where $t \ge t_{0}$ for some  $t_{0} \in \mathbb{R}_{+}$.  
\end{defn}
\begin{defn} \label{assumption_1}
	An input $u(t)$ is bounded if $\exists$ $m \in \mathbb{R}_{+}$ that satisfies $||u(t)|| \le m$, $\forall t \ge 0$.
\end{defn}
\begin{defn} \label{assumption_2}
	The uncertainty in model states $\Delta_{x}(t)$, and the output $\Delta_{y}(t)$ is bounded if $\exists$ $\delta_{m}, \delta_{n} \in \mathbb{R}_{+}$ that satisfies $\sup_{t\ge 0}||{\Delta_{x}(t)}|| \le \delta_{m}$ and $\sup_{t\ge 0}||{\Delta_{y}(t)}|| \le \delta_{n}$.
\end{defn}
\begin{property} \label{property_1}
	$\forall$ $a, b \in \mathbb{R}$ if $\sgn(a) = - \sgn(b)$ always holds true, then $\sgn(a-b) = \sgn(a) = -\sgn(b)$.
\end{property}
\begin{property} \label{property_2}
	$\forall$ $a \in \mathbb{R}$, $a = \sgn(a) |a|$.
\end{property}

\section{ Electrochemical Battery Model: ESPM Governing Equations}  \label{sec:Echem_model}
%
\textcolor{black}{In this work, a temperature-dependent ESPM, published in the literature \cite{tanim2015temperature}, has been used to simulate the transport of lithium ions in the solid and electrolyte phase, and predict the battery voltage response.} %
The ESPM governing equations describing the mass and charge transport, with a radial domain of $ r \in [0,R_{j}]$ and Cartesian domain of $l \in [0, L]$ (where $L = L_n +L_s+L_p$), are spelled out in Table \ref{ESPM_equations}.
%

\begin{table*}[!]
	\renewcommand{\arraystretch}{2}
	\caption{\textcolor{black}{Governing Equations of a temperature-dependent ESPM  \cite{tanim2015temperature, ramadass2004development}}}
	\centering
	\label{ESPM_equations}
	\centering
	\resizebox{17cm}{!}{		
		\begin{tabular}{l l l}
			\hline\hline \\[-6mm]
			\multicolumn{1}{c}{Variable} &
			\multicolumn{1}{c}{Equation} & \multicolumn{1}{c}{Boundary Condition}  \\[1.6ex] \hline
			{\pbox{20cm}{Mass transport in \\ solid phase}} & \textcolor{black}{$\dfrac {\partial c_{s,j}} {\partial t}\left(r, t\right) = D_{s,j}\left(T\right) \left [\dfrac{2} {r} \dfrac {\partial c_{s,j}} {\partial r}\left(r, t\right) + \dfrac {\partial{^2c_{s,j}}} {\partial{r^2}}\left(r, t\right) \right] $} & {\pbox{20cm}{\textcolor{black}{$\dfrac {\partial c_{s,j}} {\partial r}\left(r, t\right) \Bigr|_{\substack{r=0}} = 0$} \\ \textcolor{black}{$\quad \dfrac {\partial c_{s,j}} {\partial r} \left(r, t\right)\Bigr|_{\substack{r=R_j}}=  \dfrac{\pm I_{batt}\left(t\right)} {F  a_{s,j}  D_{s,j}\left(T\right) A  L_j}$}}} \\ [5mm]
			{\pbox{20cm}{Mass transport in \\ electrolyte phase}} & \textcolor{black}{$\epsilon_{e,j} \dfrac {\partial c_{e}} {\partial t}\left(c_{e}, t\right) = \dfrac { \partial} {\partial l} \left(D^{eff}_{e}\left(c_{e}, T\right) \dfrac{\partial c_{e}}{\partial l}\left(l, t\right) \right) + \left (1-t^{+}_{0} \right) \dfrac{\pm I_{batt}\left(t\right)}{F  A  L_j}$} & {\pbox{30cm}{\textcolor{black}{$\dfrac {\partial c_{e}} {\partial l}\Bigr|_{\substack{l=0}} = \dfrac {\partial c_{e}} {\partial l}\Bigr|_{\substack{l=L}}=  0$} \\
					\textcolor{black}{$D^{eff}_{e,n}\left(c_{e,n}, T\right) \left(\dfrac {\partial c_{e,n}} {\partial l}\left(l, t\right) \right)\Bigr|_{\substack{l=L_{n}}} = D^{eff}_{e,s}\left(c_{e,s}, T\right) \left( \dfrac {\partial c_{e,s}} {\partial l}\left(l, t\right) \right)\Bigr|_{\substack{l=L_{n}}}$} \\
					\textcolor{black}{$D^{eff}_{e,s}\left(c_{e,s}, T\right) \left(\dfrac {\partial c_{e,s}} {\partial l} \left(l, t\right) \right)\Bigr|_{\substack{l=L_{n}+L_{s}}} = D^{eff}_{e,p}\left(c_{e,p}, T\right) \left( \dfrac {\partial c_{e,s}} {\partial l} \left(l, t\right) \right)\Bigr|_{\substack{l=L_{n}+L_{s}}}$}}} \\ [5mm]
			{\pbox{20cm}{Charge transport in \\ electrolyte phase}} & \textcolor{black}{$\kappa^{eff}(c_{e}, T) \dfrac {\partial^{2} \phi_{e}} {\partial l^2}\left(l\right) + \dfrac{2 R_{g} T \kappa^{eff}(c_{e}, T) \left(1-t^{+}_{0} \right)}{F} \dfrac {\partial^2 \ln{c_{e}}} {\partial l^2}\left(l, t\right) + \dfrac{\pm I_{batt}\left(t\right)}{A L_j} = 0$} &   {\pbox{30cm}{\textcolor{black}{$\dfrac {\partial \phi_{e}} {\partial l} (l)\Bigr|_{\substack{l=0}} = \dfrac {\partial \phi_{e}} {\partial l} (l)\Bigr|_{\substack{l=L}}=  0$}}} \\ [5mm]
			\hline
		\end{tabular}
	}
\end{table*}

\begin{figure}
	\centering
	\includegraphics[width=\linewidth]{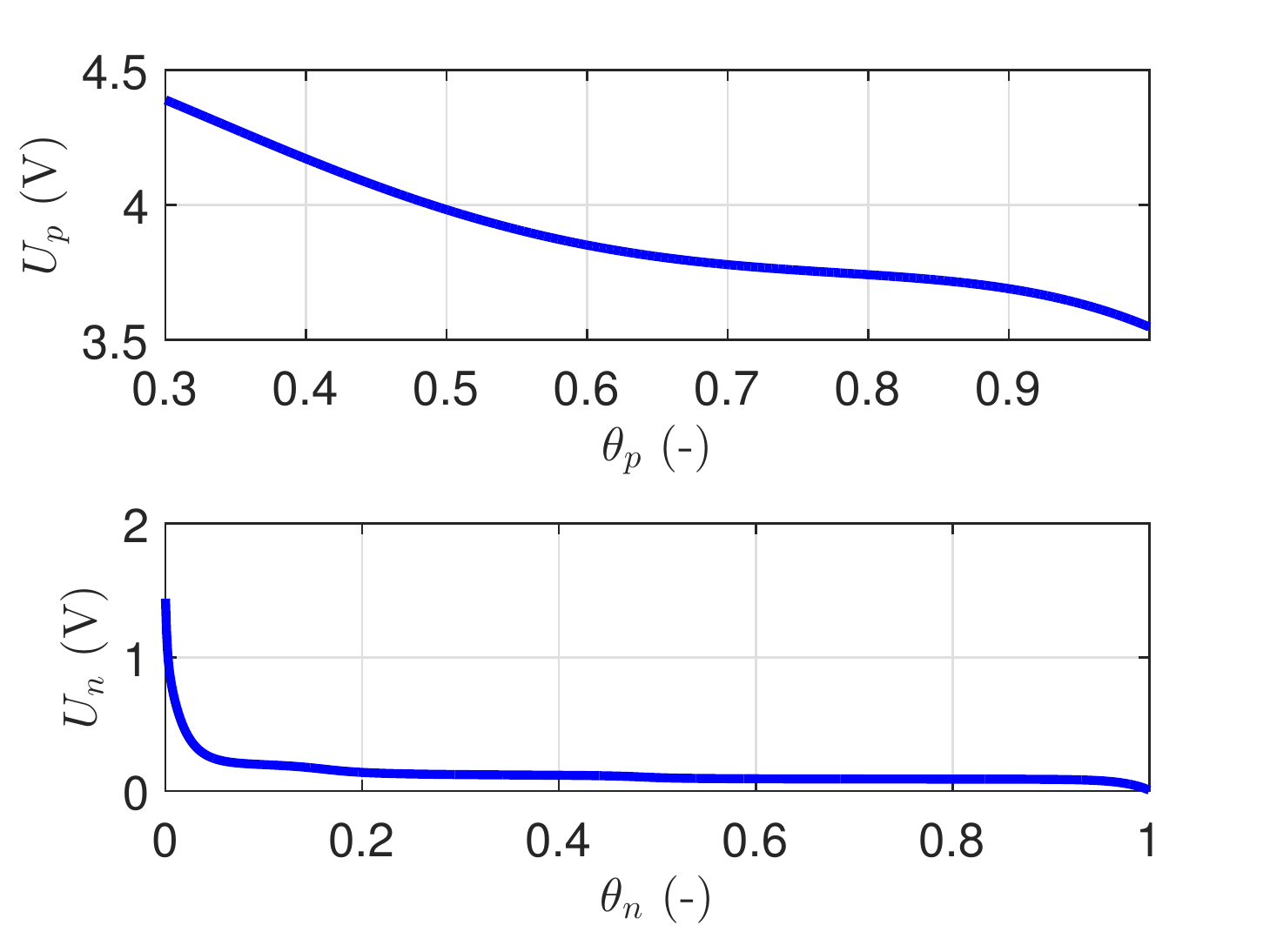}
	\caption{ The OCP of Nickel Manganese Cobalt (NMC) cathode and Graphite anode cell at $25^{o}C$.}
	\label{fig:OCP_Curves}
\end{figure}

The terminal voltage predicted by the ESPM battery model is the potential difference between cathode and anode, given by 
\begin{align} \label{eq:volt_eq_final}
	V\left(t\right) =& \left[U_p \left(c_{s,p,surf}, T \right) + \eta_p \left(c_{s,p,surf}, T, I_{batt} \right) \right]- \\ \nonumber
	&\left[U_n \left(c_{s,n,surf}, T \right) + \eta_n \left(c_{s,n,surf}, T, I_{batt} \right) \right] + \\ \nonumber
	& \dfrac{2 R_{g} T \nu(T)}{F} \ln \dfrac{c_{e}(L)}{c_e(0)} - I_{batt}\left(t\right) R_{e,0} - \\ \nonumber &I_{batt}\left(t\right) R_{l},
\end{align}
where $R_{e,0}$ is the electrolyte resistance expressed as \cite{di2010lithium}
\textcolor{black}{
\begin{align} \label{eq:electrolyte_resistance}
	R_{e,0} =& \dfrac{1}{2 A} \left( \dfrac{L_n}{\kappa^{eff}_n\left(c_{e}, T\right)} + \dfrac{2  L_{s}}{\kappa^{eff}_{s}\left(c_{e}, T\right)}+ \dfrac{ L_{p}}{\kappa^{eff}_{p}\left(c_{e}, T\right)} \right),
\end{align}
where the effective transport parameters in the electrolyte phase take tortuosity into account through a Bruggeman's relationship, to give $\kappa^{eff}_{j}(c_{e}, T) = \kappa(c_{e}, T) \epsilon_{e,j}^{1.5}$. Similar relationship holds true for the effective diffusion in electrolyte phase, appearing in Table \ref{ESPM_equations} as $D^{eff}_{e,j}(c_{e}, T) = D_{e}(c_{e}, T) \epsilon_{e,j}^{1.5}$.} %
Further, the Open Circuit Potential (OCP) of each electrode, shown in Fig~\ref{fig:OCP_Curves} \cite{tanim2015temperature}, is a function of the stoichiometry ratio, $\theta_{j}$, of the respective electrode, which is related to the surface concentration as $\theta_{j} = c_{s,j,surf}/c_{s,j,max}$, and the electro-active surface area of each electrode is defined as $a_{s,j} = 3\epsilon_{j}/R_{j}$.

\textcolor{black}{
Moreover, the dependence of the model parameters on temperature, $T$, is collated in Table \ref{ESPM_temperature_equations}.}
\begin{table}[!t]
	\renewcommand{\arraystretch}{2}
	\caption{\textcolor{black}{Temperature dependent ESPM parameters \cite{tanim2015temperature}}}
	\centering
	\label{ESPM_temperature_equations}
	\centering
	\resizebox{8cm}{!}{		
		\begin{tabular}{l l}
			\hline\hline \\[-7mm]
			\multicolumn{1}{c}{Variable} &
			\multicolumn{1}{c}{Equation} \\[0.6ex] \hline
			{\pbox{20cm}{\textcolor{black}{Solid phase} \\ \textcolor{black}{diffusion}}} & \textcolor{black}{$D_{s,j}(T)= D_{s,j,ref} \cdot exp \left[ \frac{-E_{a,D,j}}{R_g} \left(\frac{1}{T} - \frac{1}{T_{ref}}\right)\right]$} \\ [3mm]
			{\pbox{20cm}{\textcolor{black}{Reaction} \\ \textcolor{black}{rate}}} & \textcolor{black}{$k_{j}(T)= k_{j,ref} \cdot exp \left[ \frac{-E_{a,k,j}}{R_g} \left(\frac{1}{T} - \frac{1}{T_{ref}}\right)\right]$} \\ [3mm]
			{\pbox{40cm}{\textcolor{black}{Electrolyte} \\ \textcolor{black}{phase} \\ \textcolor{black}{diffusion}}}& \textcolor{black}{$D_{e}(c_{e},T)= 10^{- \left[4.43 + \dfrac{54}{T-(229+c_{e})} + 0.22c_{e}\right]} $}\\ [6mm]
			{\pbox{60cm}{\textcolor{black}{Electrolyte} \\ \textcolor{black}{phase} \\ \textcolor{black}{conductivity}}} & {\pbox{60cm}{\textcolor{black}{$\kappa(T)= c_{e} [(-10.5 + 0.074T - 6.96 \times 10^{-5}T^2 ) +$}  \\ \textcolor{black}{$c_{e} (0.668 - 0.0178T - 2.8 \times 10^{-5}T^2) + $} \\ \textcolor{black}{$c_{e}^2 (0.494 - 8.86 \times 10^{-4}T )] ^2 $}}}\\ [6mm]
			{\pbox{60cm}{\textcolor{black}{Diffusional} \\ \textcolor{black}{conductivity} \\ \textcolor{black}{(empirical)}}} & {\pbox{60cm}{\textcolor{black}{$\nu(T) = 0.601 -0.24c^{0.5}_{e} + 0.982 [1-0.0052(T-293)]c^{1.5}_{e}$}}}\\  [6mm]		
			{\pbox{60cm}{\textcolor{black}{Open Circuit} \\ \textcolor{black}{Potential}}} & {\pbox{60cm}{\textcolor{black}{$U_{j}(c_{s,j,surf},T) = U_{j}(c_{s,j,surf},T_{ref}) + $}  \\  \textcolor{black}{$\frac {\partial U_j}{\partial T}(T-T_{ref})$ }}}\\ 
			[0.4ex]
			\hline\hline
		\end{tabular}
	}
\end{table}	

%
%
%

\subsection{SEI Layer: Capacity and Power Fade Relationship}
The electrochemical instability of the electrolyte at lower potentials, typically observed at the anode, results in electrolyte decomposition forming a passive film layer on the anode, known as the SEI layer \cite{vetter2005ageing}. 
The SEI layer continues to grow, with time and usage, resulting in capacity fade and power fade of the battery. The SEI layer dynamics depends on the side reaction current density, and is described as \cite{ramadass2004development}
\begin{align}
	\label{eq:SEI_equation}
	\dfrac{\mathrm{d}L_{sei}}{\mathrm{d}t} = -\dfrac{i_{s} M_{sei}}{2F\rho_{sei}}, 
\end{align}
with initial value of $L_{sei}(0) = L_{sei,0}$ as the nominal SEI layer thickness at the Beginning of Life (BOL) of the cell after few cycles. 

\textit{Capacity fade}:
Capacity fade is defined as the decrease in the discharge capacity of the battery over time. In this work, any decrease in capacity is due to the loss of cyclable lithium ions consumed by the SEI layer. 
This allows to relate the capacity loss of the battery to the side reaction current density, and also, from \eqref{eq:SEI_equation}, relates capacity loss to the rate of SEI layer growth as \cite{prada2012simplified}
\begin{align}
	\label{eq:cap_is}
	\frac{\mathrm{d}Q}{dt} &= \dfrac{i_s  a_{s,n}  A  L_n}{3600} \\
	\dfrac{\mathrm{d}Q}{\mathrm{d}t} &= -\dfrac{\mathrm{d}L_{sei}}{\mathrm{d}t} \dfrac{2 F  \rho_{sei}  a_{s,n}  A  L_{n} }{3600M_{sei}}\label{eq:cap_SEI},
\end{align}
with initial value of $Q(0) = Q_{0}$, expressed in $Ah$, as the nominal cell capacity at the BOL. 

\textit{Power fade}:
Power fade is defined as an increase in internal resistance of the battery that results in a decrease in the power that can be delivered to the load. Under the assumption that SEI layer growth is the dominant aging mechanism, power fade is characterized through a combinatorial increase in the (a) SEI layer resistance $R_{sei}$, and (b) electrolyte resistance $R_{e}$ \cite{prada2012simplified}. If $R_{pf}$ denotes the resistance increase that represents power fade, any change in $R_{pf}$ is due to $R_{sei}$ and $R_{e}$ given by 
\begin{equation}
	\label{eq:power_fade_res}
	\dfrac{\mathrm{d}R_{pf}}{\mathrm{d}t} = \dfrac{\mathrm{d}R_{sei}}{\mathrm{d}t} + \dfrac{\mathrm{d}R_{e}}{\mathrm{d}t},
\end{equation}
with the initial value of $R_{pf}(0) = 0 \Omega $, which increases as battery ages. Integrating above equation with respect to time gives
\begin{align}
	\label{eq:power_fade_res_integrate}
	R_{pf}(t)  =& R_{sei}(t) - R_{sei}(0) + R_{e}(t) - R_{e}(0),
\end{align}
where $R_{e}(0)$ is the electrolyte resistance at BOL given in \eqref{eq:electrolyte_resistance}, hence $R_{e,0} = R_{e}(0)$.
As the SEI layer grows in thickness, the change in SEI layer resistance is related to loss in capacity using \eqref{eq:cap_SEI} as given by \cite{prada2012simplified}
\begin{align}
\label{eq:SEI_Rsei}
\dfrac{\mathrm{d}R_{sei}(t)}{\mathrm{d}t} &= \dfrac{\mathrm{d}L_{sei}\left(t\right)}{\mathrm{d}t} \dfrac{1}{a_{s,n}AL_{n}\kappa_{sei}}\\
& = -\dfrac{\mathrm{d}Q\left(t\right)}{\mathrm{d}t} \dfrac{3600M_{sei}}{2F\rho_{sei}A^{2}a_{s,n}^{2}L_{n}^{2}\kappa_{sei}}\label{eq:SEI_Rsei_2}.
\end{align}
Integrating above equation with respect to time gives
\begin{align}
	\label{eq:SEI_Rsei_3}
	R_{sei}(t) - R_{sei}(0) = - \dfrac{3600\left(Q(t) - Q_{0} \right)M_{sei}}{2F\rho_{sei}A^{2}a_{s,n}^{2}L_{n}^{2}\kappa_{sei}}. 
\end{align}
Moreover, as SEI layer continues to grow, it begins to penetrate the pores of the negative electrode restricting the accessible electroactive surface area of the electrode \cite{sikha2004effect}. This results in a modified negative electrode porosity that varies with aging given by \cite{prada2012simplified} 
\begin{align}
	\label{eq:porosity}
	\epsilon_{e,n}(t) = 1 - \epsilon_{n} \left(1+\dfrac{3L_{sei}(t)}{R_{n}}\right)-\epsilon_{n,f}.
\end{align}
The decreasing porosity affects the averaged transport properties (ionic conductivity and diffusion coefficient) in the electrolyte phase. The reduced effective ionic conductivity increases the resistance offered to lithium transport in the electrolyte phase. Integrating \eqref{eq:cap_SEI} from time $0$ to $t$, with initial conditions $L_{sei,0}, Q_{0}$, substituting the result in \eqref{eq:electrolyte_resistance}, and updating it with the modified porosity from \eqref{eq:porosity} gives the expression in \eqref{eq:Re_third_eq}.
Substituting \eqref{eq:SEI_Rsei_3} and \eqref{eq:Re_third_eq} in \eqref{eq:power_fade_res_integrate} relates the power fade resistance, $R_{pf}$ at any time $t$ to capacity $Q(t)$, as shown in \eqref{eq:Power_fade_resistance}.
\begin{figure*}
	\begin{align}
	\label{eq:Re_third_eq}
	R_{e}\left(t\right) =& \dfrac{1}{2A} \Bigg[\dfrac{L_{n}}{\kappa_{n}\Bigg(1 - \epsilon_{n} \bigg(1+\dfrac{3}{R_{s,n}}} \dfrac{}{\bigg(L_{sei,0}- \dfrac{ 3600\left(Q\left(t\right) - Q_{0}\right) M_{Sei}}{2FAL_{n}a_{s,n}\rho_{sei}}\bigg) \bigg) -\epsilon_{n,f}\Bigg)^{1.5}} + \dfrac{2L_{s}}{\kappa^{eff}_{s}} + \dfrac{L_{p}}{\kappa^{eff}_{p}}  \Bigg].
	\end{align}
	\begin{align}
	\label{eq:Power_fade_resistance}
	R_{pf}\left(t\right) =& \dfrac{1}{2A} \Bigg[\dfrac{L_{n}}{\kappa_{n}\Bigg(1 - \epsilon_{n} \bigg(1+\dfrac{3}{R_{s,n}}} \dfrac{}{\bigg(L_{sei,0}- \dfrac{ 3600\left(Q\left(t\right) - Q_{0}\right) M_{Sei}}{2FAL_{n}a_{s,n}\rho_{sei}}\bigg) \bigg) -\epsilon_{n,f}\Bigg)^{1.5}} + \dfrac{2L_{s}}{\kappa^{eff}_{s}} + \dfrac{L_{p}}{\kappa^{eff}_{p}}  \Bigg] - R_{e,0}  - \\ \nonumber
	& \dfrac{3600\left(Q(t) - Q_{0} \right)M_{sei}}{2F\rho_{sei}A^{2}a_{s,n}^{2}L_{n}^{2}\kappa_{sei}}.
	\end{align}
\end{figure*}
The novelty of deriving $R_{pf}$ this way is in establishing a tangible dependence between power fade resistance, $R_{pf}$, and capacity fade, $Q-Q_{0}$. The main characteristics of this derived relationship are (i) to supplement the conventional ESPM by including an aging-dependent term to reflect the SEI induced degradation in the form of voltage loss (the term in red box, below) in the cell terminal voltage equation as given below 
\begin{align}
	\label{eq:aging_voltage_eq}
	V &= \left[U_p \left(c_{s,p,surf}, T \right) + \eta_p \left(c_{s,p,surf}, T, I_{batt} \right) \right]- \\ \nonumber
	&\left[U_n \left(c_{s,n,surf}, T \right) + \eta_n \left(c_{s,n,surf}, T, I_{batt} \right) \right] + \\ \nonumber
	&\dfrac{2 R_{g} T \left(1-t^{+}_{0} \right) \nu(T) }{F}  \ln \dfrac{c_{e}(L)}{c_e(0)} - I_{batt}\left(t\right)R_{e,0} -  \\ \nonumber
	&  I_{batt}\left(t\right) R_{l} - \cfbox{red}{$I_{batt}\left(t\right)R_{pf}\left(t\right)$}, 
\end{align}
and (ii) to formulate the ESPM voltage equation, as in \eqref{eq:aging_voltage_eq}, in a fashion that lends itself for the estimation of available cell capacity ($Q$) by being able to monitor the voltage loss or the parameter representing the power fade resistance ($R_{pf}$).
\subsection{State Space Representation}
The Partial Differential Equations (PDEs) describing the mass transport in solid and electrolyte phase, given in Table \ref{ESPM_equations}, are spatially discretized using the Finite Difference Method (FDM) to obtain a system of coupled Ordinary Differential Equations (ODEs) that can be cast into a state space formulation. 
Moreover, the slowly varying battery capacity $Q(t)$, over its entire lifetime, is considered as a dynamic state and augmented to the state vector in order to formulate a state estimation problem. Since the capacity is a slowly varying variable, the dynamics of cell capacity in real-time is approximated as $\dot{Q} = 0$.
The system of ODEs and the aging-enhanced nonlinear terminal voltage equation in \eqref{eq:aging_voltage_eq} are formulated into a general state space form. Let $x =\left[x_{1}, x_{2}, x_{3}, x_{4} \right]^{T} \in \mathbb{R}^{(2N + M - 1) \times 1}$ be the state vector, $u=I_{batt}$ be the input current, and $y=V$ be the output voltage of the model. The state variables represent lithium concentration in cathode, anode, cell capacity, and lithium concentration in electrolyte,  ${x}_{1} = \left[c_{s,p,1}, c_{s,p,2}, \dots, c_{s,p,N} \right]^{T} $, 
${x}_{2} = \left[c_{s,n,1}, c_{s,n,2}, \dots, c_{s,n,N} \right]^{T} $, ${x}_{3} = Q$,
${x}_{4} =\left[c_{e,1}, c_{e,2}, \dots, c_{e,M-2} \right]^{T}$. Moreover, the surface concentration in both electrodes is given as $c_{s,j,surf}=Cc_{s,j}$, respectively, where $C$ is the output distribution vector. 
Then the state space formulation of ESPM is given by 
\begin{align}
	\label{eq:state_space}
	\dot{x}_{1}\left(t\right) &= A_{11}(T) x_{1}\left(t\right) + B_{1} u\left(t\right) \nonumber \\
	\dot{x}_{2}\left(t\right) &= \theta_{1}(T) \bar{A}_{22} x_{2}\left(t\right) + B_{2} u\left(t\right) \nonumber \\
	\dot{x}_{3}\left(t\right) &= 0 \\
	\dot{x}_{4}\left(t\right) &= f_{e} \left(x_{4}, T, u \right) \nonumber \\
	y(t) &= h_{1}(x_{1,N}, T, u) - h_{2}(x_{2,N}, T, u) - h_{3}(x_{3})u + \nonumber \\ \nonumber
	& h_{4}(x_{4}, T, u) - R_{l}u + \left(x_{3} - Q_{0}\right)\theta_{2}u,
\end{align}
where nonlinearities in the terminal voltage equation, and parameters are \begin{align}
		h_{1}(x_{1,N}, T, u) &= \left[U_p \left(c_{s,p,surf}, T\right) + \eta_p \left(c_{s,p,surf}, T, I_{batt} \right) \right], \nonumber \\ \nonumber
		h_{2}(x_{2,N}, T, u) &= \big[U_n \left(c_{s,n,surf}, T \right) + \\ \nonumber
		&\eta_n \left(c_{s,n,surf}, T, I_{batt} \right) \big], \nonumber \\ \nonumber
		h_{3}(x_{3}) &= R_{e}(t), \nonumber \\ \nonumber
		h_{4}(x_{4}, u)&= \dfrac{2 R_{g} T \nu(T) }{F} \ln \dfrac{c_{e}(L)}{c_e(0)}, \\ \nonumber
		\theta_{1}(T) &= D_{s,n}(T), \nonumber \\ \nonumber		
		\theta_{2} &= \dfrac{3600M_{sei}}{2FA^{2}\rho_{sei}a_{s,n}^{2}L_{n}^{2}\kappa_{sei}},
\end{align} 
and square matrices $A_{11}(T), \bar{A}_{22} \in \mathbb{R}^{N \times N} $ are the coefficients of the concentration states in \eqref{eq:state_space}, and column vectors $B_{1}, B_{2} \in \mathbb{R}^{N \times 1}$ are coefficients of input current in \eqref{eq:state_space}, described as given below
	\begin{align}
	\label{eq:ODE_matrices}
	\textcolor{black}{A_{11}(T)} &= \textcolor{black}{\dfrac{D_{s,p}(T)}{\Delta_{r}^2}} \begin{bmatrix} -2 & 2 & 0 & \cdots & 0 & 0\\ 1/2 & -2 & 3/2 & \cdots & 0 & 0 \\ \vdots &  \vdots & \vdots & \ddots & \vdots & \vdots  \\ 0 & 0 & 0 & \cdots & 2 & -2 \end{bmatrix} \\ \nonumber 
	B_{1} &= \dfrac{2}{\Delta_r  F  a_{s,p}  A  L_{p}} \left[ \begin{array}{c} 0 \\ 0 \\ \vdots \\ \dfrac{N+1}{N} \end{array} \right]  \\ \nonumber
	\bar{A}_{22} &= \dfrac{1}{\Delta_{r}^2} \begin{bmatrix} -2 & 2 & 0 & \cdots & 0 & 0\\ 1/2 & -2 & 3/2 & \cdots & 0 & 0 \\ \vdots &  \vdots & \vdots & \ddots & \vdots & \vdots  \\ 0 & 0 & 0 & \cdots & 2 & -2 \end{bmatrix} \\ \nonumber 
	B_{2} &= \dfrac{-2}{\Delta_r  F  a_{s,n}  A  L_{n}} \left[ \begin{array}{c} 0 \\ 0 \\ \vdots \\ \dfrac{N+1}{N} \end{array} \right]. \nonumber
	\end{align}
The procedure used to identify the ESPM parameters, and validate it against experimental data is outlined in Appendix. 

\section{Adaptive Interconnected Observer} \label{sec:Adaptive_Obs}
Accurate knowledge of battery SOC/SOH using the state space model described in \eqref{eq:state_space} is attainable by estimating the following state variables: 1) lithium concentration in cathode, 2) lithium concentration in anode, and 3) total cell capacity. However, it is important to note that accurate model-based state estimation over the entire lifespan of a battery is contingent on how well the model predicts the battery response as it ages. 
Naturally, when model parameters vary with usage and time, state estimates of capacity and lithium concentration will diverge from their respective true values. 
Studies have shown that 
transport parameters such as diffusion and conductivity change with aging \cite{ramadass2004development}.  
This motivates the need for an adaptive scheme that updates the time-varying aging-sensitive parameters in real-time to ensure that model-based estimation of capacity and lithium concentration remains accurate over time. 
For this purpose, an adaptive observer capable of combined estimation of states and parameters is considered.  A sliding mode interconnected observer structure \cite{allam2018interconnected} is preferred for the implementation of the adaptive observer, primarily because it allows for the concurrent estimation of concentration in both electrodes, and by extension, estimation of electrode-specific geometrical and transport parameters, despite any inaccurate initialization in either electrode. %
The observability issues associated with estimating states from both electrodes are circumvented by having an observer for each electrode with an open loop model of the other electrode that is constantly updated with the correct estimates. More importantly, the sliding mode structure features robustness to modeling uncertainties and easier real-time on-board implementation.  In this work, the SEI layer growth is considered to be the major degradation mechanism, and hence anode diffusion coefficient $({D}_{s,n})$ and SEI layer ionic conductivity $({\kappa}_{sei})$ are chosen as the parameters of interest that are assumed to change with degradation. The changes in anode diffusion due to SEI layer are well documented \cite{ramadass2004development}, and the lowering of ionic conductivity in SEI layer is interpreted from the growing SEI layer thickness and subsequent increasing SEI layer resistance offered to the transport of lithium ions. 
Both parameters are studied to be moderately sensitive to the output voltage and hence can be estimated with a reasonable level of accuracy. It is worth pointing out that the interconnected framework can be easily extended to incorporate different degradation mechanisms that affect other parameters at either electrode, since there is a dedicated observer running for each electrode. 

\begin{figure}\centering
	\includegraphics[width=8cm,keepaspectratio]{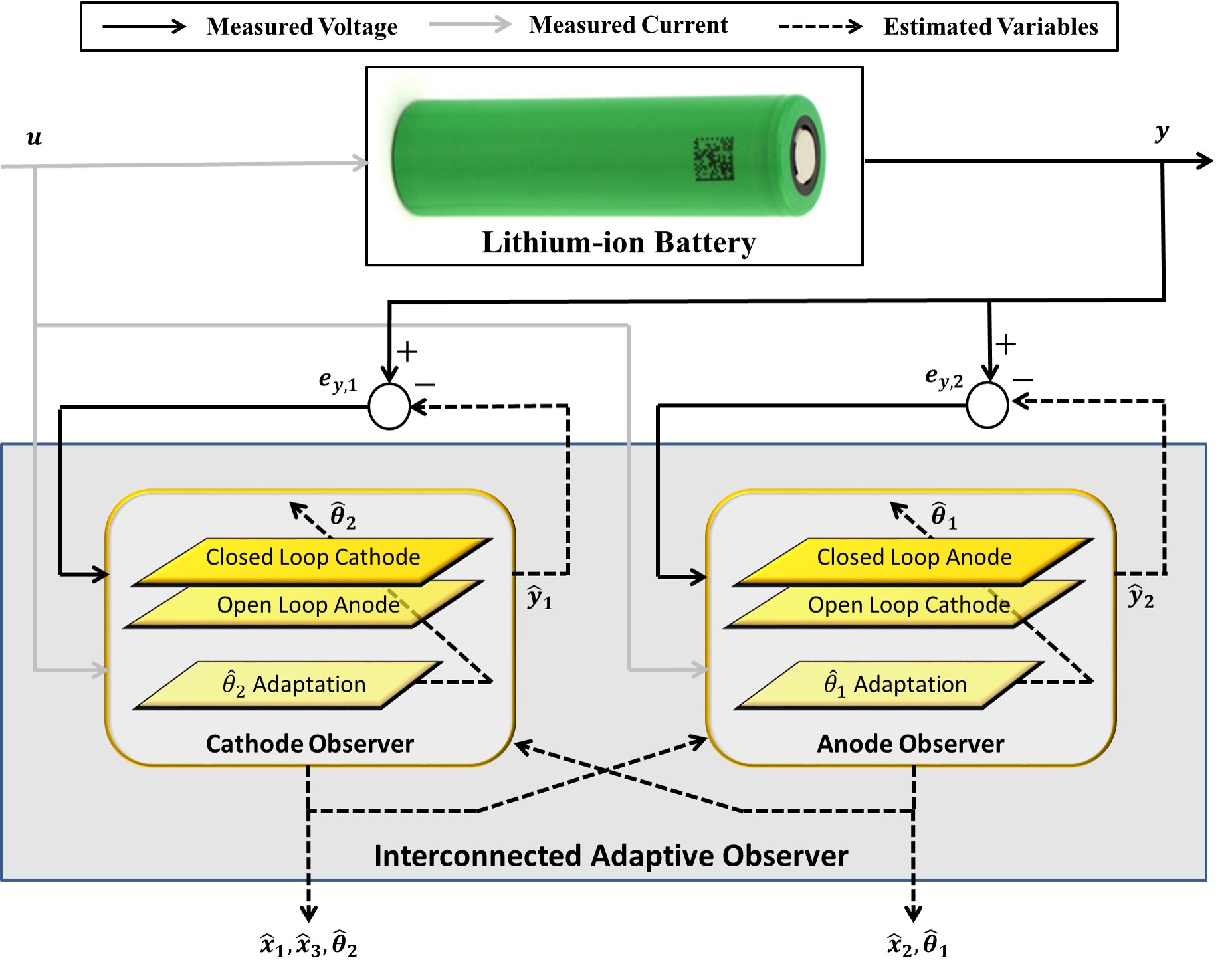}
	\caption{ The interconnected adaptive observer structure for the estimation of lithium concentration states ($\hat{x}_{1}, \hat{x}_{2}$), total cell capacity ($\hat{x}_{3}$), anode diffusion coefficient ($\hat{\theta}_{1}$), and the SEI layer ionic conductivity ($\hat{\theta}_{2}$).}
	\label{fig:interconn_adaptive}
\end{figure}

The state ($x$) and parameter ($\theta$) vectors to be estimated are
$x = [x_{1}, x_{2}, x_{3}]^{T} \in \mathbb{R}^{(2N+1) \times 1} $ and 
$\theta  = [\theta_{1}, \theta_{2}]^{T} \in \mathbb{R}^{2 \times 1}$.
The structure of the proposed adaptive interconnected observer is illustrated in Fig.~\ref{fig:interconn_adaptive}. %
The observer is fed with measured current and voltage of the battery. The \textit{cathode observer} estimates : 1) the lithium concentration in the cathode $({x}_{1})$, 2) the cell capacity $({x}_{3})$, 3) the SEI layer ionic conductivity $\kappa_{sei} (\theta_{2})$.
Whereas, the \textit{anode observer} estimates 1)the lithium concentration in the anode $({x}_{2})$, 2) the anode diffusion coefficient $D_{s,n} (\theta_{1})$. Recall that the parameter $(\kappa_{sei})$ enters the state space model in \eqref{eq:state_space} through the term $\theta_{2}$. Only the term $(\kappa_{sei})$ is unknown in $\theta_{2}$. Hence estimating $\theta_{2}$ and using the values of the remaining known parameters gives $\kappa_{sei}$.  
 The estimated state variables and parameters from one observer are fed to the other, at every step, through a bidirectional interconnection, guaranteeing each observer to converge despite incorrect initialization in states and parameters. While the convergence of the sliding mode
interconnected observer for state estimation with fixed battery model parameters is proved for a fresh cell \cite{allam2018interconnected}, the convergence for a cell whose aging-sensitive transport parameters vary slowly over time is proposed in this paper, which to the best of the authors' knowledge has never been investigated. 

\begin{remark}\label{rem:SPM_for_obs}
	For observer design, the ESPM is simplified by assuming uniform concentration in the electrolyte phase. The concentration state in the electrolyte phase $(x_{4})$ is considered to have a constant value of $1200$ $mol/m^3$ \cite{tanim2015temperature}, hence $\dot{x}_{4} = 0$. Effectively, the $M-2$ ODEs representing the electrolyte phase are eliminated and the term $\dfrac{2 R_{g} T \left(1-t^{+}_{0} \right) \nu(T)}{F} \ln \dfrac{c_{e}(L)}{c_e(0)}$ in the output voltage equation is taken to be zero. This gives an approximated SPM that is a reduced-order model suitable for observer design. This also allows comparison of the SPM-based observer estimates with the higher order ESPM. 
	\textcolor{black}{
	Finally, the combined uncertain state space representation of the SPM is given as
	\begin{align}
		\label{eq:ODE_matrices2}
		\left[ \begin{array}{c} \dot{x}_{1} \\ \dot{x}_{2} \\ \dot{x}_{3} \end{array} \right] =& \begin{bmatrix} A_{11}(T) & 0_{N \times N} & 0_{N \times 1} & \\ 0_{N \times N} & \theta_{1}(T) \bar{A}_{22} & 0_{N \times 1} \\ 0 & 0_{1 \times N} & 0_{1 \times N} \end{bmatrix} \left[ \begin{array}{c} x_{1} \\ x_{2} \\ x_{3} \end{array} \right] + \nonumber \\  
		&\left[ \begin{array}{c} B_{1} \\ B_{2} \\ 0 \end{array} \right] u   + \nonumber \\
		& \begin{bmatrix} I_{N} & 0_{N \times N} & 0_{N \times 1} & \\ 0_{N \times N} & I_{N} & 0_{N \times 1} \\ 0_{1 \times N} & 0_{1 \times N} & 0 \end{bmatrix} \left[ \begin{array}{c} \Delta_{x_{1}} \\ \Delta_{x_{2}} \\ 0 \end{array} \right] \\ \nonumber
		\left[ \begin{array}{c} \dot{\theta}_{1} \\ \dot{\theta}_{2} \end{array} \right] =& \left[ \begin{array}{c} 0 \\ 0 \end{array} \right] \\ \nonumber
		y =& h_{1}(x_{1,N}, T, u) - h_{2} (x_{2,N}, T, u)- R_{l}u - \\ \nonumber
		&h_{3}(x_{3})u + \left(x_{3} - Q_{0}\right)\theta_{2}u + \Delta_{y},
	\end{align}
}
	where $\Delta_{x_{1}}$, $\Delta_{x_{2}}$ $\in \mathbb{R}^{N \times 1}$, and $\Delta_{y} \in \mathbb{R}$ are the modeling uncertainties introduced in the states and output due to neglecting the concentration dynamics in the electrolyte phase. %
	\textcolor{black}{Henceforth, the dependence on cell temperature $T$ is dropped only in the notations, for the ease of presentation. The cell temperature information is assumed to be known via temperature sensors, and the model states and output are updated accordingly based on the temperature value.}
	

\end{remark}

\begin{theorem}
	For the dynamical state space representation of SPM with known and bounded uncertainties in states ( $\Delta_{x_{1}}$, $\Delta_{x_{2}}$) and output ($\Delta_{y}$), given in \eqref{eq:ODE_matrices2}, if 
	\begin{enumerate}
		\item \label{thm_condition_21} there exists functions $h_{1} \left( x_{1,N}, u \right), h_{2} \left( x_{2,N}, u \right)  : \mathbb{R} \times \mathbb{R} \times \mathbb{R} \rightarrow \mathbb{R} $ which are Lipschitz in $x_{1,N}$ and $x_{2,N}$, respectively, and uniformly in $u$;%
		\item \label{thm_condition_2} there exists a function $h_{3} \left( x_{3} \right) : \mathbb{R} \rightarrow \mathbb{R} $ Lipschitz in $x_{3}$;%
		%
		%
		%
		\item \label{thm_condition_5} the input $u$ is bounded and satisfies the property of persistence of excitation as per Definition \ref{defn:pers_excite}; 
		\item \label{thm_condition_4} the output uncertainty $\Delta_{y}$ is related to the error in capacity estimate, through $\psi \in \mathbb{R}$, as follows
		\begin{equation}
		\Delta_{y} = \psi e_{3} u,
		\end{equation}
		\item \label{thm_condition_6} and the parameters are adapted, with tuning parameters $k_{1}, k_{2}$, according to
		\begin{align}
			\begin{cases}
				\dot{\hat{\theta}}_{1} &= \dfrac{C \bar{A}_{22}\hat{x}_{2} \sgn \left(e_{y_{2}}\right) |\tilde{h}_{2}|}{\gamma_{n,2} k_{1}} \\
				{\dot{\hat{\theta}}}_{2} &= \dfrac{CG_{1} \left(\hat{{x}}_{3} - Q_{0}\right)u\sgn(e_{y_{1}})|\tilde{h}_{1} - \tilde{h}_{2}|}{k_{2}\gamma_{p,2}} 	
			\end{cases}
		\end{align}		
	\end{enumerate}
	then the adaptive interconnected observer, consisting of a \textit{cathode observer} formulated as
	\begin{align}
		\label{eq:ss_obsv_p}
		\dot{\hat{x}}_{1} =& A_{11} \hat{x}_{1} + B_{1}u + G_{1} \left(y-\hat{y}_{1} \right) + G_{v1} \sgn \left(y-\hat{y}_{1} \right) \nonumber \\
		\dot{\hat{x}}_{2,ol}  =& \hat{\theta}_{1} \bar{A}_{22} \hat{x}_{2} + B_{2}u  \nonumber \\
		\dot{\hat{x}}_{3} =& G_{3} \left(y-\hat{y}_{1} \right) u \nonumber \\
		\hat{y}_{1} =& h_{1} \left( \hat{x}_{1,N}, u \right) - h_{2} \left( \hat{x}_{2,N,ol}, u \right) - R_{l}u - \\ \nonumber
		&h_{3}(\hat{x}_{3})u  + \left(\hat{x}_{3} - Q_{0}\right)\hat{\theta}_{2}u, \nonumber
	\end{align}
	and an \textit{anode observer} formulated as
	\begin{align}
		\label{eq:ss_obsv_n}
		\dot{\hat{x}}_{1,ol}  =& A_{11} \hat{x}_{1} + B_{1}u  \nonumber \\
		\dot{\hat{x}}_{2} =& \hat{\theta}_{1} \bar{A}_{22} \hat{x}_{2} + B_{2}u + G_{2} \left(y-\hat{y}_{2} \right) +G_{v2} \sgn \left(y-\hat{y}_{2} \right) \nonumber \\
		\hat{y}_{2} =& h_{1} \left( \hat{x}_{1,N,ol}, u \right) - h_{2} \left( \hat{x}_{2,N}, u \right) - R_{l}u -  \\ \nonumber
		&h_{3}(\hat{x}_{3})u + \left(\hat{x}_{3} - Q_{0}\right)\hat{\theta}_{2}u, \nonumber
	\end{align}
	is practically stable, i.e. the state and parameter estimates converge to a bounded error ball as $t \rightarrow \infty$.	
\end{theorem}

\begin{remark}
	In \eqref{eq:ss_obsv_p} and \eqref{eq:ss_obsv_n}, the subscript $ol$ stands for open loop model state variables, $G_{1} \in \mathbb{R}^{N \times 1}_{-}, G_{2} \in \mathbb{R}^{N \times 1}_{+}$, $G_3 \in \mathbb{R}$  are constant linear observer gains, $ G_{v1}, G_{v2} \in \mathbb{R}^{N \times 1}$ are variable structure gains, introduced to improve robustness against uncertainties, with discontinuous injection terms defined as
	\begin{align}
		\sgn \left(y-\hat{y}_{i} \right) = 
		\begin{cases}
			1,& \text{if } y-\hat{y}_{i} > 0\\
			0, &\text{if }  y-\hat{y}_{i} = 0\\
			-1, & \text{if }  y-\hat{y}_{i} < 0
		\end{cases} \nonumber
		i=1, 2.
	\end{align}
	%
\end{remark}
\begin{remark} \label{remark:surface_err_C_matrix}
	The error in the surface concentration of cathode $(e_{1,N})$ is related to the entire error vector of cathode concentration via the output distribution vector as
	\begin{align}
		e_{1,N} =& Ce_{1}.
	\end{align} 
	The same holds true for anode: 	$e_{2,N} = Ce_{2}$.	
\end{remark}	

\begin{figure}[!t]\centering
	\includegraphics[width=\linewidth]{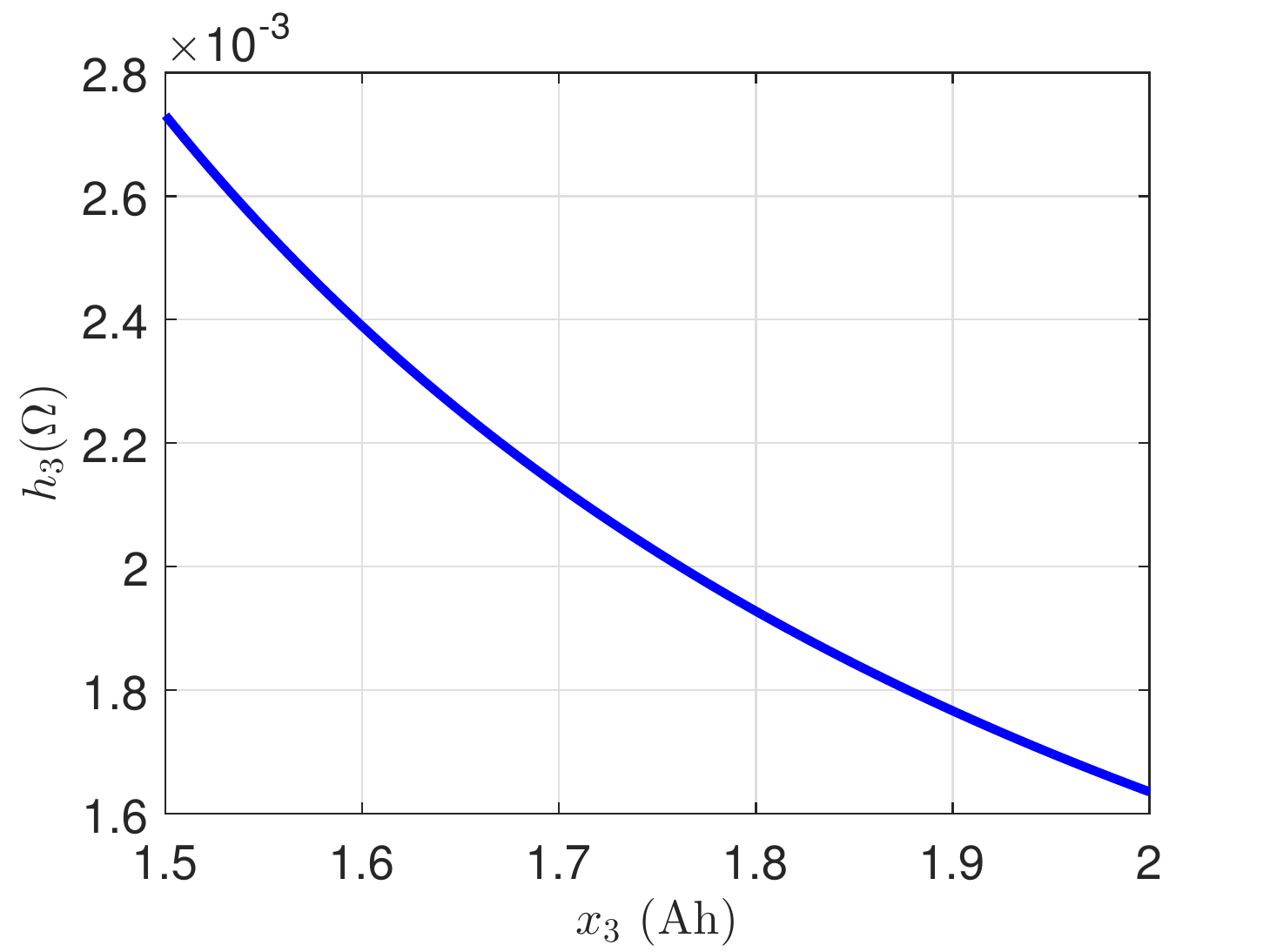}
	\caption{Function $h_{3}$ plotted against $x_{3}$ with nominal parameter values.} \label{fig:h3_lipschitz}
\end{figure}

\begin{remark}\label{remark:sign_convention}
During battery operation, it is important to understand that the lithium cycling between the two electrodes results in the concentration in one electrode to increase, while the concentration in the other electrode decreases. This understanding is exploited in the observer formulation and initialization. Consider the stoichiometric window of anode to be $\theta_{n,100\%}$ and $\theta_{n,0\%}$ corresponding to fully charged (100$\%$ SOC) and fully discharged (0$\%$ SOC) cell, and likewise, the cathode stoichiometric window as $\theta_{p,100\%}$ and $\theta_{p,0\%}$   corresponding to fully charged and fully discharged cell. If we discharge the cell from a fully charged status, the stoichiometry of anode will start from $\theta_{n,100\%}$ and move towards $\theta_{n,0\%}$, where $\theta_{n,100\%}>\theta_{n,0\%}$. On the other hand, the stoichiometry of cathode will start from $\theta_{p,100\%}$   and move towards $\theta_{p,0\%}$, where $\theta_{p,100\%}<\theta_{p,0\%}$. This is because the concentration in anode will deplete as the concentration in cathode increases.
For instance if the true SOC is 100$\%$ and the cell is initialized with an error of 10$\%$ (i.e. SOC = 90$\%$; note that SOC = 110$\%$ is not a feasible initialization because it is not physically possible), then this error is introduced into the concentration state variables of the observer in terms of initial stoichiometry values of anode and cathode as  $\theta_{n,initial}$ and $\theta_{p,initial}$, respectively. From the above understanding, we are aware that these initial values will always have to lie within the stoichiometric windows of the respective electrode for feasibility. This leads to
\begin{align}
\begin{cases}
\theta_{n,100\%}>\theta_{n,initial}>\theta_{n,0\%} \\  \theta_{p,100\%}<\theta_{p,initial}<\theta_{p,0\%}
\end{cases}
\end{align} 
Physically, there cannot be a value of $\theta_{p,initial}>\theta_{p,100\%}$ that can satisfy or correspond to SOC = 90$\%$.
Hence, we can write that the sign of the error at the anode stoichiometry is opposite to that of the sign of the error at the cathode stoichiometry, given as 
\begin{align}
\sgn(\theta_{n,100\%}-\theta_{n,initial} )=-\sgn(\theta_{p,100\%}-\theta_{p,initial}).
\end{align}
This relation holds true for the surface stoichiometry or the surface concentration of the respective electrodes, which gives 
\begin{align}
\sgn \left(x_{1,N} - \hat{x}_{1,N} \right) = - \sgn \left(x_{2,N} - \hat{x}_{2,N} \right).
\end{align}
\end{remark}
\begin{remark} \label{remark:monotonous_functions}
	Functions $h_{1} \left( x_{1,N}, u\right)$, $h_{2} \left( x_{2,N}, u \right)$ as shown in Fig.~\ref{fig:OCP_Curves}, and $h_{3}\left(x_{3}\right)$ as shown in Fig.~\ref{fig:h3_lipschitz}, are Lipschitz in $x_{1,N}, x_{2,N}, $ and $x_{3}$, respectively. Moreover, the functions are strictly monotonically decreasing functions, and their gradients are bounded as follows
	\begin{align}
	\label{eq:lip_ineq}
		\begin{cases}
			-\gamma_{p,1} &\le \dfrac{\partial h_{1}}{\partial x_{1,N}} \le -\gamma_{p,2} \\
			-\gamma_{n,1} &\le \dfrac{\partial h_{2}}{\partial x_{2,N}} \le -\gamma_{n,2} \\
			-\alpha_{Q,1} &\le \dfrac{\partial h_{3}}{\partial x_{3}} \le -\alpha_{Q,2},		
		\end{cases}
	\end{align} 
	where $\gamma_{p,1}, \gamma_{p,2}, \gamma_{n,1}, \gamma_{n,2}, \alpha_{Q,1}, \alpha_{Q,2} \in \mathbb{R}_{+}$.	
\end{remark}	

\begin{lemma} \label{lemma:monotonous_functions}
The inequalities
\begin{align} 
\begin{cases}
-e^{T}_{1}G_{1}\tilde{h}_{1} \le e^{T}_{1}G_{1}\gamma_{p,2}Ce_{1} \\  -e^{T}_{2}G_{2}\tilde{h}_{2} \le e^{T}_{2}G_{2}\gamma_{n,2}Ce_{2},
\end{cases}
\end{align}
hold true regardless of the sign of the errors $e_{1}$, $e_{2}$.
\\

Rewriting the first expression from Remark \ref{remark:monotonous_functions} in \eqref{eq:lip_ineq} as
\begin{align}
\label{eq:lipschitz_bounds}
-\gamma_{p,1} \le \dfrac{h_{1}(x_{1,N})-h_{1}(\hat{x}_{1,N})}{x_{1,N}-\hat{x}_{1,N}} \le -\gamma_{p,2}.
\end{align}
Using Remark \ref{remark:surface_err_C_matrix}, and considering the scenario where $e_{1}<0$, which implicitly means  $e_{1,N}<0$, and multiplying by $e_{1,N}$ on both sides of \eqref{eq:lipschitz_bounds}, causes the inequalities to change giving
\begin{align}
\label{eq:lipschitz_bounds_neg1}
-\gamma_{p,1}e_{1,N} \ge h_{1}(x_{1,N})-h_{1}(\hat{x}_{1,N}) \ge -\gamma_{p,2}e_{1,N} \nonumber \\
-\gamma_{p,1}Ce_{1} \ge \tilde{h}_{1} \ge -\gamma_{p,2}Ce_{1}.
\end{align}
Since $G_{1}\in\mathbb{R}^{N \times 1}_{-}$, the product $-e^{T}_{1}G_{1}$ will always be negative ($-e^{T}_{1}G_{1}<0$). Multiplying $-e^{T}_{1}G_{1}$on both sides of \eqref{eq:lipschitz_bounds_neg1} causes the inequality sign to change leading to 
\begin{align}
\label{eq:lipschitz_bounds_neg2}
e^{T}_{1}G_{1}\gamma_{p,1}Ce_{1} \le -e^{T}_{1}G_{1}\tilde{h}_{1} \le e^{T}_{1}G_{1}\gamma_{p,2}Ce_{1}.
\end{align}
Further, consider the scenario where $e_{1}>0$, which implicitly means  $e_{1,N}>0$, and multiplying by $e_{1,N}$ on both sides of \eqref{eq:lipschitz_bounds} gives
\begin{align}
\label{eq:lipschitz_bounds_pos1}
-\gamma_{p,1}e_{1,N} \le h_{1}(x_{1,N})-h_{1}(\hat{x}_{1,N}) \le -\gamma_{p,2}e_{1,N} \nonumber \\
-\gamma_{p,1}Ce_{1} \le \tilde{h}_{1} \le -\gamma_{p,2}Ce_{1}.
\end{align}
In this case, the product $-e^{T}_{1}G_{1}$ will always be positive ($-e^{T}_{1}G_{1}>0$). Multiplying $-e^{T}_{1}G_{1}$on both sides of \eqref{eq:lipschitz_bounds_pos1} gives
\begin{align}
\label{eq:lipschitz_bounds_pos2}
e^{T}_{1}G_{1}\gamma_{p,1}Ce_{1} \le -e^{T}_{1}G_{1}\tilde{h}_{1} \le e^{T}_{1}G_{1}\gamma_{p,2}Ce_{1}.
\end{align}
From \eqref{eq:lipschitz_bounds_neg2} and \eqref{eq:lipschitz_bounds_pos2}, it is clear that the following inequality \begin{align}
-e^{T}_{1}G_{1}\tilde{h}_{1} \le e^{T}_{1}G_{1}\gamma_{p,2}Ce_{1}
\end{align} 
always hold true, irrespective of the sign of the estimation error $e_{1}$. 
And likewise $-e^{T}_{2}G_{2}\tilde{h}_{2} \le e^{T}_{2}G_{2}\gamma_{n,2}Ce_{2}$ for the second expression in \eqref{eq:lip_ineq}. 

\end{lemma}
\begin{proof}
	Define the errors for the cathode observer as
	\begin{align}
		\label{eq:error_define_cathode}
		\begin{cases}
			{e}_{1} &= x_{1} - \hat{x}_{1} \\
			{e}_{2,ol} &= x_{2} - \hat{x}_{2,ol} \\
			{e}_{3} &= x_{3} - \hat{x}_{3} \\
			{e}_{\theta_{2}} &= \theta_{2} - \hat{\theta}_{2}, \\
		\end{cases}
	\end{align}
	and for the anode observer as 
	\begin{align}
		\label{eq:error_define_anode}
		\begin{cases}
			{e}_{1,ol} &= x_{1} - \hat{x}_{1,ol} \\
			{e}_{2} &= x_{2} - \hat{x}_{2} \\
			{e}_{\theta_{1}} &= \theta_{1} - \hat{\theta}_{1}. \\
		\end{cases}
	\end{align}
	From \eqref{eq:ss_obsv_p} and \eqref{eq:ss_obsv_n} the error dynamics for the state estimation are written as
	\begin{align}
		\begin{cases}
			\dot{e}_{1} &= A_{11} e_{1}-  G_{1} \left(y-\hat{y}_{1} \right)  - G_{v1} \sgn \left(y-\hat{y}_{1} \right) + \Delta_{x_{1}} \\ 
			\dot{e}_{2} &= {\theta}_{1} \bar{A}_{22} e_{2} + e_{\theta_1} \bar{A}_{22} \hat{x}_{2} - G_{2} \left(y-\hat{y}_{2} \right) - \\ 
			&G_{v2} \sgn \left(y-\hat{y}_{2} \right) + \Delta_{x_{2}}\\ 
			\dot{e}_{3} &= -  G_{3} \left(y-\hat{y}_{1} \right) u. \nonumber
		\end{cases}
	\end{align}
	For parameters that are slowly varying, the following assumption is made: $ {\dot{\theta}}_{1} = {\dot{\theta}}_{2} = 0 $, hence the error dynamics for parameter estimation are given by
	\begin{align}
		\label{eq:ss_obsv_param_err}
		\begin{cases}
			\dot{e}_{\theta_{1}} &= {\dot{\theta}}_{1} - {\dot{\hat{\theta}}}_{1} = - {\dot{\hat{\theta}}}_{1} \\
			\dot{e}_{\theta_{2}} &= {\dot{\theta}}_{2} - {\dot{\hat{\theta}}}_{2} = - {\dot{\hat{\theta}}}_{2}.
		\end{cases}
	\end{align}

	Further, the output error of the cathode and anode observer is defined as
	\begin{align}
		e_{y_{1}} = y - \hat{y}_{1} =&  \left[ h_{1} \left( x_{1,N}, u \right) -  h_{1} \left( \hat{x}_{1,N}, u \right) \right]- \nonumber \\ \nonumber
		&\left[  h_{2} \left( x_{2,N}, u \right) - h_{2} \left( \hat{x}_{2,N,ol}, u \right) \right] - \\ \nonumber
		&\left[ h_{3}({x}_{3})u - h_{3}(\hat{x}_{3})u \right] + \\ \nonumber 
		&\left[\left({x}_{3} - Q_{0}\right){\theta}_{2}u - \left(\hat{x}_{3} - Q_{0}\right)\hat{\theta}_{2}u \right] + \Delta_{y}.
	\end{align}
	Defining $\tilde{h}_{1} =  h_{1} \left( x_{1,N}, u \right) -  h_{1} \left( \hat{x}_{1,N}, u \right), \\
	\tilde{h}_{2} =  h_{2} \left( x_{2,N}, u \right) -  h_{2} \left( \hat{x}_{2,N}, u \right)$, and \\
	$\tilde{h}_{3} =  h_{3} \left( x_{3}\right) -  h_{3} \left( \hat{x}_{3} \right)$, gives
	%
	\begin{align} \label{eq:cathode_output_error}
		e_{y_{1}} =& \tilde{h}_{1} - \tilde{h}_{2,ol} - \tilde{h}_{3}u + \theta_{2}e_{3}u \\ \nonumber
		&+ \left(\hat{x}_{3} - Q_{0}\right)e_{\theta_{2}}u + \Delta_{y}, 
	\end{align}
	and  similarly, the anode output error is described as 
	\begin{align}\label{eq:anode_output_error}
		e_{y_{2}} = y - \hat{y}_{2} =& \tilde{h}_{1,ol} - \tilde{h}_{2} - \tilde{h}_{3}u + \theta_{2}e_{3}u \nonumber \\
		& + \left(\hat{x}_{3} - Q_{0}\right)e_{\theta_{2}}u + \Delta_{y}.
	\end{align}
	Let $V_{O}$ be the composite Lyapunov function for the interconnected observers given by
	\begin{equation}
		\label{eq:lyap}
		V_{O}\left(t\right) = V_{1}\left(t\right) + V_{2}\left(t\right),
	\end{equation}	
	where $V_{1}\left(t\right)$ and $V_{2}\left(t\right)$ are the candidate Lyapunov functions for the cathode and anode observer. %
	\textcolor{black}{
	 It is worth mentioning that the stability of individual cathode and anode observers may not guarantee the stability of the overall interconnected observer. To that end, the bidirectional information exchange between the two individual observers is taken into consideration to provide the conditions for practical stability for the whole interconnected observer. \\%
	 The Lyapunov functions for the cathode and anode observer are  defined as	
	\begin{equation}
		\label{eq:lyap_cathode}
		V_{1}\left(t\right) = \frac{1}{2} e^{T}_{1} e_{1} + \frac{1}{2} e^{2}_{3} +  \frac{1}{2}k_{2}e^{2}_{\theta_{2}}, 
	\end{equation}	
	and
	\begin{equation}
		\label{eq:lyap_anode}
		V_{2}\left(t\right) = \frac{1}{2} e^{T}_{2} e_{2} + \frac{1}{2} k_{1}e^{2}_{\theta_{1}}.
	\end{equation}
	The candidate functions in \eqref{eq:lyap_cathode} and \eqref{eq:lyap_anode} are analyzed separately, one at a time, albeit taking into account the information exchange (state variable update) from the adjacent connected observer. 
}
Taking the derivative of \eqref{eq:lyap_cathode} with respect to time, and substituting $e_{y,1} = y - \hat{y}_{1}$ from \eqref{eq:cathode_output_error} yields
	\begin{align}\label{eq:cathode_obs_eq}
		\dot{V}_{1} =& e^{T}_{1} \dot{e}_{1} + e_{3}\dot{e}_{3} + k_{2} e_{\theta_{2}} \dot{e}_{\theta_{2}}  \nonumber \\ \nonumber
		=& e^{T}_{1} A_{11} e_{1} - e^{T}_{1} G_{1}\tilde{h}_{1} + e^{T}_{1} G_{1} \tilde{h}_{2,ol} + e^{T}_{1} G_{1}\tilde{h}_{3} u - \\ \nonumber
		&  e^{T}_{1} G_{1}\theta_{2}e_{3}u - e^{T}_{1} G_{1} \left(\hat{x}_{3} - Q_{0}\right)e_{\theta_{2}}u - e^{T}_{1} G_{1}\Delta_{y} + \\ 
		&  e^{T}_{1} \Delta_{x_{1}}  - e^{T}_{1} G_{v1} \sgn \left(y-\hat{y}_{1}\right)  - e_{3}G_{3}e_{y_{1}}u  \\ \nonumber
		& - k_{2}e_{\theta_{2}} {\dot{\hat{\theta}}}_{2}.
		%
	\end{align}
	Likewise, for the anode observer, taking derivative of \eqref{eq:lyap_anode} with respect to time, and substituting $e_{y,2} = y - \hat{y}_{2}$ from \eqref{eq:anode_output_error} yields
	\begin{align}\label{eq:anode_obs_eq}
		\dot{V}_{2} =&	e^{T}_{2}\theta_{1}\bar{A}_{22} e_{2} + e^{T}_{2}e_{\theta_{1}} \bar{A}_{22} \hat{x}_{2} - e^{T}_{2}G_{2}\tilde{h}_{1,ol}  + \\ \nonumber
		& e^{T}_{2}G_{2}\tilde{h}_{2} + e^{T}_{2}G_{2}\tilde{h}_{3} u - e^{T}_{2}G_{2}\theta_{2}e_{3}u - \\ \nonumber
		&e^{T}_{2}G_{2}\left(\hat{x}_{3} - Q_{0}\right)e_{\theta_{2}}u -e^{T}_{2}G_{2} \Delta_{y} + e^{T}_{2}\Delta_{x_{2}}  - \\ \nonumber
		& e^{T}_{2}G_{v2} \sgn \left(y-\hat{y}_{2}\right) - k_{1} e_{\theta_{1}} {\dot{\hat{\theta}}}_{1}. 
		%
		%
		%
	\end{align}
	Combining \eqref{eq:cathode_obs_eq}, \eqref{eq:anode_obs_eq}, and then grouping related terms, gives
	\begin{align} \label{eq:final_lyap_combined}
		\dot{V}_{O} =& e^{T}_{1} A_{11} e_{1} - e^{T}_{1} G_{1}\tilde{h}_{1} + e^{T}_{1} G_{1} \tilde{h}_{2,ol} - \\ \nonumber
		& e^{T}_{1}G_{v1} \sgn \left(y-\hat{y}_{1}\right) + e^{T}_{1}\Delta_{x_{1}} + \\ \nonumber
		& e^{T}_{2}\theta_{1}\bar{A}_{22} e_{2} - e^{T}_{2}G_{2}\tilde{h}_{1,ol}  +  e^{T}_{2}G_{2}\tilde{h}_{2} - \\ \nonumber
		& e^{T}_{2}G_{v2} \sgn \left(y-\hat{y}_{2}\right) + e^{T}_{2}\Delta_{x_{2}} + \\ \nonumber
		&  e^{T}_{1}G_{1}\tilde{h}_{3}u - e^{T}_{1}G_{1}\theta_{2}{e}_{3}u - e^{T}_{1}G_{1}\Delta_{y}  - e_{3}G_{3}e_{y_{1}}u + \\ \nonumber
		& e^{T}_{2}G_{2}\tilde{h}_{3}u - e^{T}_{2}G_{2}\theta_{2}{e}_{3}u - e^{T}_{2}G_{2} \Delta_{y} - \\ \nonumber
		& e^{T}_{1}G_{1} \left(\hat{x}_{3} - Q_{0}\right)e_{\theta_{2}}u - k_{2}e_{\theta_{2}} {\dot{\hat{\theta}}}_{2} - \\  \nonumber
		&e^{T}_{2}G_{2}\left(\hat{x}_{3} - Q_{0}\right)e_{\theta_{2}}u + \\ \nonumber
		& e^{T}_{2} e_{\theta_{1}} \bar{A}_{22}  \hat{x}_{2} -  k_{1} e_{\theta_{1}} {\dot{\hat{\theta}}}_{1}.
	\end{align}
	In \eqref{eq:final_lyap_combined}, the terms are grouped as per the state or parameter error they are related to, and denoted as follows: 
	\begin{enumerate}		
\item $\dot{V}_{c}$: first five terms are related to the cathode concentration estimation error,
\item $\dot{V}_{a}$: next five terms are related to anode concentration estimation error,
\item $\dot{V}_{q}$: followed by seven terms related to capacity estimation error,
\item $\dot{V}_{\kappa}$: next three terms for the SEI layer ionic conductivity estimation error,
\item $\dot{V}_{D_{s}}$: final two terms for the anode diffusion coefficient error.
\end{enumerate} 
	%
	
	%
	%
	
	Consider the terms denoting cathode concentration estimation error $\dot{V}_{c}$,
	\begin{align}
		\dot{V}_{c} = \dot{V}_{c,1} + \dot{V}_{c,2},
	\end{align}
	where,
	\begin{align}
		\begin{cases}
			\dot{V}_{c,1} &= e^{T}_{1} A_{11} e_{1} - e^{T}_{1} G_{1}\tilde{h}_{1}, \\
			\dot{V}_{c,2} &=  e^{T}_{1}G_{1}\tilde{h}_{2} - e^{T}_{1}G_{v,1}\sgn\left(y_{1}-\hat{y}_{1}\right) + e^{T}_{1} \Delta_{x_{1}}.
		\end{cases}
	\end{align}
	Consider $\dot{V}_{c,1}$,
	and using Lemma \ref{lemma:monotonous_functions}, gives
	\begin{align}
		\dot{V}_{c,1} &\le  e^{T}_{1} A_{11} e_{1} + e^{T}_{1} G_{1}\gamma_{p,2}Ce_{1}. \\ \nonumber
		\dot{V}_{c,1} &\le  e^{T}_{1} \left( A_{11} + e^{T} G_{1}C\gamma_{p,2} \right) e_{1}.
	\end{align}	
	Since the lithium concentration of a single electrode is observable from the voltage equation \cite{bartlett2016electrochemical}, there exists a gain $G_{1}\in\mathbb{R}_{-}^{N \times 1}$ that places all the eigenvalues of $A_{11} + G_{1}\gamma_{p,2}C$ in the left half plane, making $\dot{V}_{c,1}$ negative definite.\\
	\\
	Now consider $\dot{V}_{c,2}$,
	\begin{align}
		\dot{V}_{c,2} &=  e^{T}_{1}G_{1}\tilde{h}_{2,ol} - e^{T}_{1}G_{v,1}\sgn\left(y_{1}-\hat{y}_{1}\right) + e^{T}_{1} \Delta_{x_{1}}.
	\end{align}
		Note that in a lithium-ion battery cell, the lithium ions are transported from one electrode to another. Hence, as the lithium concentration in anode increases, the concentration in cathode decreases, and vice-versa. 
	
From Remark \ref{remark:monotonous_functions}, the nonlinear functions $h_{1}(x_{1,N})$ and $h_{2}(x_{2,N})$ are strictly monotonically decreasing functions in $x_{1,N}$ and $x_{2,N}$, respectively. Further, using Remark \ref{remark:sign_convention}, it can be inferred that	
	\begin{align}
		\begin{cases}
			\sgn \left(x_{1,N} - \hat{x}_{1,N} \right) &= - \sgn \left(x_{2,N} - \hat{x}_{2,N} \right),  \\
			\sgn \left(h_{1}(x_{1,N}) - h_{1}(\hat{x}_{1,N}) \right) &= - \sgn \left(x_{1,N} - \hat{x}_{1,N} \right),  \\
			\sgn \left(h_{2}(x_{2,N}) - h_{2}(\hat{x}_{2,N})\right) &= - \sgn \left(x_{2,N} - \hat{x}_{2,N} \right).
		\end{cases}
	\end{align}
	It follows that
	\begin{align} 
		\begin{cases}
			\label{eq:sign_h_function}
			\sgn(e_{1,N}) =& \sgn(e_{1}) = \sgn(\tilde{h}_{2}),  \\
			\sgn(e_{2,N}) =& \sgn(e_{2}) = \sgn(\tilde{h}_{1}),  \\
			\sgn(\tilde{h}_{1}) =& -\sgn(\tilde{h}_{2}).
		\end{cases}
	\end{align} 
	
	Consider the output voltage equation given in \eqref{eq:cathode_output_error}, Since $\tilde{h}_{1}$ and $\tilde{h}_{2,ol}$ will always have opposing signs, from \eqref{eq:sign_h_function}, the difference $\tilde{h}_{1}- \tilde{h}_{2,ol}$ will always add up in magnitude and have the sign same as the sign of the first element in the difference, which in this case is $\tilde{h}_{1}$. Clearly, the magnitude and sign of the cathode observer output voltage error will be dominated by the difference  $\tilde{h}_{1}- \tilde{h}_{2,ol}$ compared to the remaining aging-related terms in \eqref{eq:cathode_output_error}, which are : $\theta_{2}e_{3}u + \left(\hat{x}_{3} - Q_{0}\right)e_{\theta_{2}}u + \Delta_{y}$. Therefore, it can be written that
	\begin{align} 
		\sgn(e_{y_{1}}) = \sgn(y_{1} - \hat{y}_{1}) &= \sgn \left(\tilde{h}_{1}- \tilde{h}_{2,ol} \right) \\ \nonumber
		&= - \sgn \left(\tilde{h}_{2,ol}- \tilde{h}_{1} \right). \nonumber
	\end{align}
	Again, due to the opposing signs of $\tilde{h}_{1}$ and $\tilde{h}_{2,ol}$, the sign of the difference will be always same as the first element in the difference from Property \ref{property_1}, giving 
	\begin{align} 
		\sgn(e_{y_{1}}) =  \sgn(y_{1} - \hat{y}_{1}) = - \sgn \left(\tilde{h}_{2,ol} \right).
	\end{align}
	Using this knowledge in $\dot{V}_{c,2}$, choosing gain $G_{v1}$ be related to gain $G_{1}$ through a scalar relationship given by  $G_{v1} = -\beta_{1}G_{1}$ where $\beta_{1} \in \mathbb{R}_{+}$, and using Property \ref{property_2}, gives
	\begin{align}\label{eq:Vc2_resolve}
		\dot{V}_{c,2} &=  e^{T}_{1}G_{1}\tilde{h}_{2,ol} - \beta_{1} e^{T}_{1}G_{1}\sgn \left(\tilde{h}_{2,ol} \right) + e^{T}_{1} \Delta_{x_{1}} \\ \nonumber
		&=  e^{T}_{1}G_{1}\tilde{h}_{2,ol} \left(1 - \dfrac{ \beta_{1}}{\lvert \tilde{h}_{2,ol} \rvert} \right) +  e^{T}_{1} \Delta_{x_{1}}.	
	\end{align}
	Since $G_{1}$ is always negative, and $\sgn(e_{1}) = \sgn(\tilde{h}_{2,ol})$, the following condition always holds true, irrespective of the sign of the elements of $e_{1}$
	\begin{align}
		\sgn \left(e^{T}_{1}G_{1}\tilde{h}_{2,ol} \right) = -1.
	\end{align}
	The above  relationship is rewritten as
	\begin{align}
		e^{T}_{1}G_{1}\tilde{h}_{2,ol} = - \lvert  e^{T}_{1}G_{1}\tilde{h}_{2,ol} \rvert.
	\end{align}
	Substituting back in \eqref{eq:Vc2_resolve}, gives
	\begin{align} \label{eq:Vc2_resolve_2}
		\dot{V}_{c,2} &=  - \lvert  e^{T}_{1}G_{1}\tilde{h}_{2,ol} \rvert \left(1 - \dfrac{\beta_{1} }{\lvert \tilde{h}_{2,ol} \rvert} \right) +  e^{T}_{1} \Delta_{x_{1}}. 
	\end{align}
	The following condition on $\beta_{1}$ ensures $\dot{V}_{c,2} \le 0$
	\begin{align} \label{eq:beta_1}
		\beta_{1}&\le \left( \dfrac{-e^{T}_{1} \Delta_{x_{1}}}{\lvert  e^{T}_{1}G_{1}\tilde{h}_{2,ol} \rvert} + 1 \right) \lvert \tilde{h}_{2,ol} \rvert.
	\end{align}
	This ensures that both $\dot{V}_{c,1}$ and $\dot{V}_{c,2}$ are negative definite, resulting in $\dot{V}_{c}$ to decay to a bounded error ball whose radius is determined by the modeling uncertainty $\Delta_{x_{1}}$. 
	%
	%
	%
	%
	
	Further, the anode concentration error terms $\dot{V}_{a}$ are,
	\begin{align}
		\label{eq:V_a_1}
		\dot{V}_{a} =& e^{T}_{2}\theta_{1}\bar{A}_{22} e_{2} - e^{T}_{2}G_{2}\tilde{h}_{1,ol}  +  e^{T}_{2}G_{2}\tilde{h}_{2} \\ \nonumber
		& e^{T}_{2}G_{v2} \sgn \left(y-\hat{y}_{2}\right) + e^{T}_{2}\Delta_{x_{2}}
	\end{align}
	Likewise to the aforementioned proof for cathode terms in $\dot{V}_{c}$, there exists a gain $G_{2}\in\mathbb{R}_{+}^{N \times 1}$ that places all the eigenvalues of $\theta_{1}\bar{A}_{2} + G_{2}\gamma_{n,2}C$ in the left half plane, making it negative definite. %
	Moreover, gain $G_{v2}$ is chosen to be related to gain $G_{2}$ through a scalar relationship given by  $G_{v2} = -\beta_{2}G_{2}$ where $\beta_{2} \in \mathbb{R}_{+}$. Finally, if the below given condition for $\beta_{2}$ is satisfied (which is derived in similar fashion as done above for $\beta_{1}$)  
	\begin{align} \label{eq:beta_2}
		\beta_{2}&\le \left(\dfrac{-e^{T}_{2} \Delta_{x_{2}}}{\lvert  e^{T}_{2}G_{2}\tilde{h}_{1,ol} \rvert} + 1 \right) \lvert \tilde{h}_{1,ol} \rvert,
	\end{align}
	then ${V}_{a}$ converges to a ball of radius bounded by $\Delta_{x_{2}}$. A conservative approach is undertaken to tune the values for $\beta_{1}$ and $\beta_{2}$, by selecting  values for $e_{1}, e_{2}, \tilde{h}_{1,ol}$ and $\tilde{h}_{2,ol}$ that relate to acceptable initial errors.
	%
	%
	%
	%
	%
	%
	%
	%
	
	Consider the capacity estimation error related terms $\dot{V}_{q}$,
	\begin{align}
\dot{V}_{q} =& e^{T}_{1}G_{1}\tilde{h}_{3}u - e^{T}_{1}G_{1}\theta_{2}{e}_{3}u - e^{T}_{1}G_{1}\Delta_{y}  - e_{3}G_{3}e_{y_{1}}u + \\ \nonumber
& e^{T}_{2}G_{2}\tilde{h}_{3}u - e^{T}_{2}G_{2}\theta_{2}{e}_{3}u - e^{T}_{2}G_{2} \Delta_{y}
	\end{align}
	Since $h_{3}$ is Lipschitz in $x_{3}$, using Remark \ref{remark:monotonous_functions} it can be written that $\tilde{h}_{3} \le -\alpha_{Q,2}e_{3}$. The above equation is rewritten as
	\begin{align}
		\label{eq:V_q_1}
		\dot{V}_{q} \le& - e^{T}_{1}G_{1}\alpha_{Q,2}e_{3}u -e^{T}_{1}G_{1}\theta_{2}{e}_{3}u - e^{T}_{1}G_{1}\Delta_{y}  - e_{3}G_{3}e_{y_{1}}u - \nonumber \\
		&  e^{T}_{2}G_{2}\alpha_{Q,2}e_{3}u - e^{T}_{2}G_{2}\theta_{2}{e}_{3}u - e^{T}_{2}G_{2} \Delta_{y}  \\ \nonumber		
		\le& -e^{T}_{1}G_{1}\left(\theta_{2} + \alpha_{Q,2}\right){e}_{3}u - e^{T}_{1}G_{1}\Delta_{y}  - e_{3}G_{3}e_{y_{1}}u - \nonumber \\
		& e^{T}_{2}G_{2}\left(\theta_{2} + \alpha_{Q,2}\right){e}_{3}u - e^{T}_{2}G_{2} \Delta_{y}  \\ \nonumber
		\le&  -\left(e^{T}_{1} G_{1} + e^{T}_{2} G_{2}\right) \left(\theta_{2} + \alpha_{Q,2}\right)e_{3}u -\\ \nonumber
		&  \left(e^{T}_{1} G_{1} + e^{T}_{2} G_{2}\right) \Delta_{y} - e_{3}G_{3}ue_{y_{1}}.  
	\end{align}
	Assuming $\Delta_{y} = \psi e_{3} u$ since any bounded modeling uncertainty in the output will result in an error in the estimation of capacity $(e_{3})$, under any input $u$. In other words, if there is no uncertainty in the output, i.e. if $\Delta_{y} = 0$ then there would not be an error in the capacity estimate. Rewriting \eqref{eq:V_q_1} as given below
	\begin{align}
\label{eq:V_q_2}
\dot{V}_{q} \le&  -\left(e^{T}_{1} G_{1} + e^{T}_{2} G_{2}\right) \left(\theta_{2} + \alpha_{Q,2}\right) e_{3} u -\\ \nonumber
&  \left(e^{T}_{1} G_{1} + e^{T}_{2} G_{2}\right) \psi e_{3} u - e_{3}G_{3}ue_{y_{1}} \nonumber
\end{align}	
	Upon rearranging the above equation, if gain $G_{3}$ satisfies the below relationship	
	\begin{align}
		\label{eq:g3_condition}
		G_{3} \ge & \dfrac{||\left(e^{T}_{1} G_{1} + e^{T}_{2} G_{2}\right)\left(\theta_{2} + \alpha_{Q,2} + \psi  \right)|| }{||e_{y_{1}}||},
	\end{align}
	then ${V}_{q}$ converges to a ball of radius bounded by $\psi$ and the steady state estimation errors of anode and cathode concentration states $x_{1}$ and $x_{2}$. Acceptable initial error values for $e_{1}, e_{2}$ and $e_{y,1}$ are chosen to tune the value of gain $G_{3}$.
	%
	%
	%

	For the SEI layer ionic conductivity estimation error terms,
	\begin{align}
		\dot{V}_{\kappa} =& -e^{T}_{1}G_{1} \left(\hat{x}_{3} - Q_{0}\right)e_{\theta_{2}}u - \\ \nonumber &e^{T}_{2}G_{2}\left(\hat{x}_{3} - Q_{0}\right)e_{\theta_{2}}u - k_{2} e_{\theta_{2}} {\dot{\hat{\theta}}}_{2} \\  \nonumber		
		=& -e^{T}_{1}G_{1} \left(\hat{x}_{3} - Q_{0}\right)e_{\theta_{2}}u -  e^{T}_{2}G_{2}\left(\hat{x}_{3} - Q_{0}\right)e_{\theta_{2}}u\\  \nonumber
		&-\dfrac{k_{2} e_{\theta_{2}} {\dot{\hat{\theta}}}_{2} \sgn(e_{y_{1}})}{\sgn(e_{y_{1}})}.
	\end{align}
	The estimation of SEI layer ionic conductivity is intended to begin after the lithium concentration estimates for both electrodes converge to the error ball, so that the SEI layer ionic conductivity does not show transients due to the initial error in electrode lithium concentration. This enables the assumption that $\tilde{h}_{2,ol} = \tilde{h}_{2}$, which means that the open loop model of anode in the cathode observer has been corrected and it gives the same estimate as that of the closed loop model of anode in the anode observer. It follows that $\sgn(e_{y_{1}}) = \sgn(\tilde{h}_{1} - \tilde{h}_{2,ol})$, and using Property \ref{property_2} and Remark \ref*{remark:monotonous_functions}, 
	\begin{align}
\sgn(\tilde{h}_{1} - \tilde{h}_{2}) &=  \dfrac{\tilde{h}_{1} - \tilde{h}_{2}}{|\tilde{h}_{1} - \tilde{h}_{2}|} \\
&=  \dfrac{\tilde{h}_{1}}{|\tilde{h}_{1} - \tilde{h}_{2}|} - \dfrac{\tilde{h}_{2}}{|\tilde{h}_{1} - \tilde{h}_{2}|} \\
 &\le -\dfrac{\gamma_{p,2}Ce_{1}}{|\tilde{h}_{1} - \tilde{h}_{2}|} + \dfrac{\gamma_{n,2}Ce_{2}}{|\tilde{h}_{1} - \tilde{h}_{2}|}.
	\end{align}
	
	 This leads to
	\begin{align}
		\dot{V}_{\kappa} \le& -e^{T}_{1}G_{1} \left(\hat{x}_{3} - Q_{0}\right)e_{\theta_{2}}u - e^{T}_{2}G_{2}\left(\hat{x}_{3} - Q_{0}\right)e_{\theta_{2}}u  \\  \nonumber
		&+\dfrac{k_{2} e_{\theta_{2}} {\dot{\hat{\theta}}}_{2} \gamma_{p,2}Ce_{1}}{\sgn(e_{y_{1}})|\tilde{h}_{1} - \tilde{h}_{2}|}-\dfrac{k_{2} e_{\theta_{2}} {\dot{\hat{\theta}}}_{2} \gamma_{n,2}Ce_{2}}{\sgn(e_{y_{1}})|\tilde{h}_{1} - \tilde{h}_{2}|}.
	\end{align}
	Rearranging the terms, and with the knowledge that for any scalar, $Ce_{1} = e^{T}_{1}C^{T}$ and $Ce_{2} = e^{T}_{2}C^{T}$, we have
	\begin{align}
		\dot{V}_{\kappa} \le& \bigg(-e^{T}_{1}G_{1} \left(\hat{x}_{3} - Q_{0}\right)u + \dfrac{ e^{T}_{1}C^{T}k_{2}\gamma_{p,2}{\dot{\hat{\theta}}}_{2}}{ \sgn(e_{y_{1}})|\tilde{h}_{1} - \tilde{h}_{2}|}\bigg)e_{\theta_{2}} \\  \nonumber	
		& - \bigg(e^{T}_{2}G_{2}\left(\hat{x}_{3} - Q_{0}\right)e_{\theta_{2}}u + \dfrac{  e^{T}_{1}C^{T}k_{2}\gamma_{n,2}{\dot{\hat{\theta}}}_{2}}{ \sgn(e_{y_{2}})|\tilde{h}_{1} - \tilde{h}_{2}|} \bigg) e_{\theta_{2}}.
	\end{align}	
	The terms inside the parentheses can be set to $0$, if the following two adaptation laws hold true 
	\begin{align}
		\begin{cases}
		\label{eq:theta_2_cond}
			{\dot{\hat{\theta}}}_{2} =& \dfrac{CG_{1} \left(\hat{x}_{3} - Q_{0}\right)u\sgn(e_{y_{1}})|\tilde{h}_{1} - \tilde{h}_{2}|}{k_{2}\gamma_{p,2}} \\
			{\dot{\hat{\theta}}}_{2} =& -\dfrac{CG_{2} \left(\hat{x}_{3} - Q_{0}\right)u\sgn(e_{y_{1}})|\tilde{h}_{1} - \tilde{h}_{2}|}{k_{2}\gamma_{n,2}}, \\	
		\end{cases}
	\end{align}
	which is only possible if the gains of the cathode and anode observers are chosen to satisfy the following relationship
	\begin{align}
	\label{eq:G1_G2_relation}
		\dfrac{G_{1}}{\gamma_{p,2}} = - \dfrac{G_{2}}{\gamma_{n,2}}.
	\end{align}	
	 Note that $|\tilde{h}_{1} - \tilde{h}_{2}|$ in \eqref{eq:theta_2_cond} is unknown in real-time, and hence a tolerable value is chosen. This leads to a conservative solution but ensures that ${V}_{\kappa}$ only decays to a bounded region characterized by the steady-state errors in the estimation of $x_{1}, x_{2}, x_{3}$, since estimation of $x_{1}, x_{2}, x_{3}$ only converges to their respective error balls. %
	 Further, the adaptation law for $\theta_{2}$ requires the input current $u$ to satisfy the persistence of excitation condition.
	
	Finally, for the error terms related to the anode diffusion coefficient estimation,
	\begin{align}
		\dot{V}_{D_{s}} =& e^{T}_{2}\bar{A}_{22} e_{\theta_{1}} \hat{x}_{2} -  {k_{1}e_{\theta_{1}}  \dot{\hat{\theta}}}_{1} \\ \nonumber
		=&  \left(e^{T}_{2} \bar{A}_{22} \hat{x}_{2} - \dfrac{k_{1} {\dot{\hat{\theta}}}_{1} \sgn \left(e_{y_{2}} \right)}{\sgn \left(e_{y_{2}} \right)} \right) e_{\theta_{1}}. \nonumber 
	\end{align}
	Using Property \ref{property_1} and \ref{property_2}, $\sgn \left(e_{y_{2}} \right) = -\sgn \left(\tilde{h}_{2} \right) = \dfrac{-\tilde{h}_{2}}{|\tilde{h}_{2}|}$, and knowing $\tilde{h}_{2} \le -\gamma_{n,2}Ce_{2}$ and $Ce_{2} = e^{T}_{2}C^{T}$ gives
	\begin{align}
		\dot{V}_{D_{s}} \le&  \left(e^{T}_{2} \bar{A}_{22} \hat{x}_{2} - \dfrac{e^{T}_{2}C^{T} \gamma_{n,2} k_{1} {\dot{\hat{\theta}}}_{1}}{\sgn \left(e_{y_{2}} \right)|\tilde{h}_{2}|} \right) e_{\theta_{1}}. \nonumber
	\end{align}
	Choosing the following adaptation law
	\begin{align}
			\label{eq:theta_1_cond}
		\dot{\hat{\theta}}_{1} &= \dfrac{C \bar{A}_{22}\hat{x}_{2} \sgn \left(e_{y_{2}}\right) |\tilde{h}_{2}|}{\gamma_{n,2} k_{1}},
	\end{align}
	ensures that ${V}_{D_{s}}$ decays and lies within a bounded region defined by the steady state error in estimation of $x_{2}$. Note that $\tilde{h}_{2}$ in \eqref{eq:theta_1_cond} is unknown in real-time, and hence a tolerable value is chosen resulting in a conservative approach.
	
	Combining the results from $\dot{V}_{c}, \dot{V}_{a}, \dot{V}_{q}, \dot{V}_{\kappa},$ and $\dot{V}_{D_{s}}$ yields
	\begin{align}
		\dot{V}_{O} \le& \dot{V}_{c} + \dot{V}_{a} + \dot{V}_{q} + \dot{V}_{\kappa} + \dot{V}_{D_{s}} \leq  0.
	\end{align}
	Since, ${V}_{c}, {V}_{a}, {V}_{q}, {V}_{\kappa},$ and ${V}_{D_{s}}$ converge only to a ball that is bounded by their respective modeling uncertainties and steady state errors, ${V}_{O}$ is practically stable as per Definition \ref{defn:practical_stab}. Further, the radius of the error balls can be reduced by tuning the gains $\beta_{1}, \beta_{2}, G_{3}, k_{1}$ and $k_{2}$.	 
\end{proof}

\section{Results and Discussion} \label{sec:Results_Discussion}
Two lithium-ion NMC cells (Cells $\#$ A, B) at different stages of health with distinct measured capacity values as shown in Table \ref{Cap_est}, are chosen to test the performance of the proposed interconnected observer.
Notably, cell $\#$ A is a fresh cell with a higher capacity value, whereas cell $\#$ B has been aged  under the protocol discussed in \cite{liu2017synthesis}. 
%
%
\begin{table}[!]
	\renewcommand{\arraystretch}{1.2}
	\caption{ Summary of Capacity Estimation Results for Fresh and Aged Cells for a Charge-Sustaining drive cycle (US06)}
	\centering
	\label{Cap_est}
	\resizebox{6cm}{!}{
		\begin{tabular}{l l l l}
			\hline\hline \\[-3mm]
			\multicolumn{1}{c}{Cell $\#$} &
			\multicolumn{1}{c}{\pbox{20cm}{Measured \\ Capacity (Ah)}} & 
			\multicolumn{1}{c}{\pbox{20cm}{Estimated \\ Capacity (Ah)}} &
			\multicolumn{1}{c}{\pbox{20cm}{Estimation \\ Error ($\%$)}} \\[1.6ex] \hline
			$A$  & $1.95$ & $1.94$ & $0.92$ \\
			$B$  & $1.84$ & $1.82$ & $1.65$ \\
			[0.4ex] \hline\hline
		\end{tabular}
	}
\end{table}
%
%
The experimentally measured current and voltage data of these cells, subjected to any particular drive cycle, are used as the input to the proposed interconnected adaptive observer. %
%
The estimated capacity is compared against the measured capacity of each cell. The estimation error in capacity is computed as $ Q_{err} = \frac{ \left(\hat{Q} - Q \right) }{Q} \times 100\%$, also tabulated in the last column of Table \ref{Cap_est}. On the other hand, the estimated bulk and surface concentration in both electrodes, and the aging-sensitive parameters are validated against the higher order model, ESPM,  described in Section \ref{sec:Echem_model}. %
%
\textcolor{black}{
\subsection{Observer Gains Tuning Process}
Tolerable values of errors and variables are assumed to tune the gains of the adaptive interconnected observer such that the conditions derived in Section \ref{sec:Adaptive_Obs} are satisfied. The following steps can be undertaken to tune the observer gains. %
1) Firstly, gains $G_1$ and $G_2$ are adjusted to ensure that the trajectory of the concentration estimates from an incorrect initialized value approaches the true/reference value. %
In the absence of information on the bounds on the gradients as given in \eqref{eq:G1_G2_relation}, the gains are selected by fixing $G_1$ and then tuning $G_2$ that leads to a minimum steady state error in the estimation of cathode and anode concentration. 
%
2) Next, tuning parameter $k_{1}$ is calibrated to make sure that the diffusion estimate converges to the identified diffusion coefficient of the ESPM. A tolerable value of the error $|\tilde{h}_{2}|$ in \eqref{eq:theta_1_cond} is chosen assuming the maximum error that can exist in the initial condition of solid phase concentration. In this work, the maximum initial error in the lithium concentration in both electrodes is assumed to be $45\%$ (which can be selected based on the application; for instance in a Hybrid Electric Vehicle that has a charge sustaining operation, the SOC window of operation is small and hence the initial error chosen is low, as opposed to an Electric Vehicle application where the initial error can be high), and hence the corresponding error in $|\tilde{h}_{2}|$ is considered. 
3) The gain $G_3$ and the tuning parameter $k_{2}$ are then adjusted such that the capacity and SEI layer conductivity estimates satisfy the practical stability condition in \eqref{eq:g3_condition} and \eqref{eq:theta_2_cond}. %
Again, the unknown tolerable error values in real-time are chosen by assuming the maximum initial error in the concentration of both electrodes, based on the application. 
Note that the estimation of capacity and SEI layer ionic conductivity begins after the lithium concentration estimates for both electrodes converge within their respective error ball. This is carried out to ensure that the capacity estimate does not show transients due to the high initial solid phase concentration error. Further, the capacity estimate is passed through a low pass filter to smooth out any remaining transients.
}
\subsection{Capacity estimation for cells at different stages of health}
\begin{figure}[tp]
	\centering
	{\includegraphics[width=8.5cm]{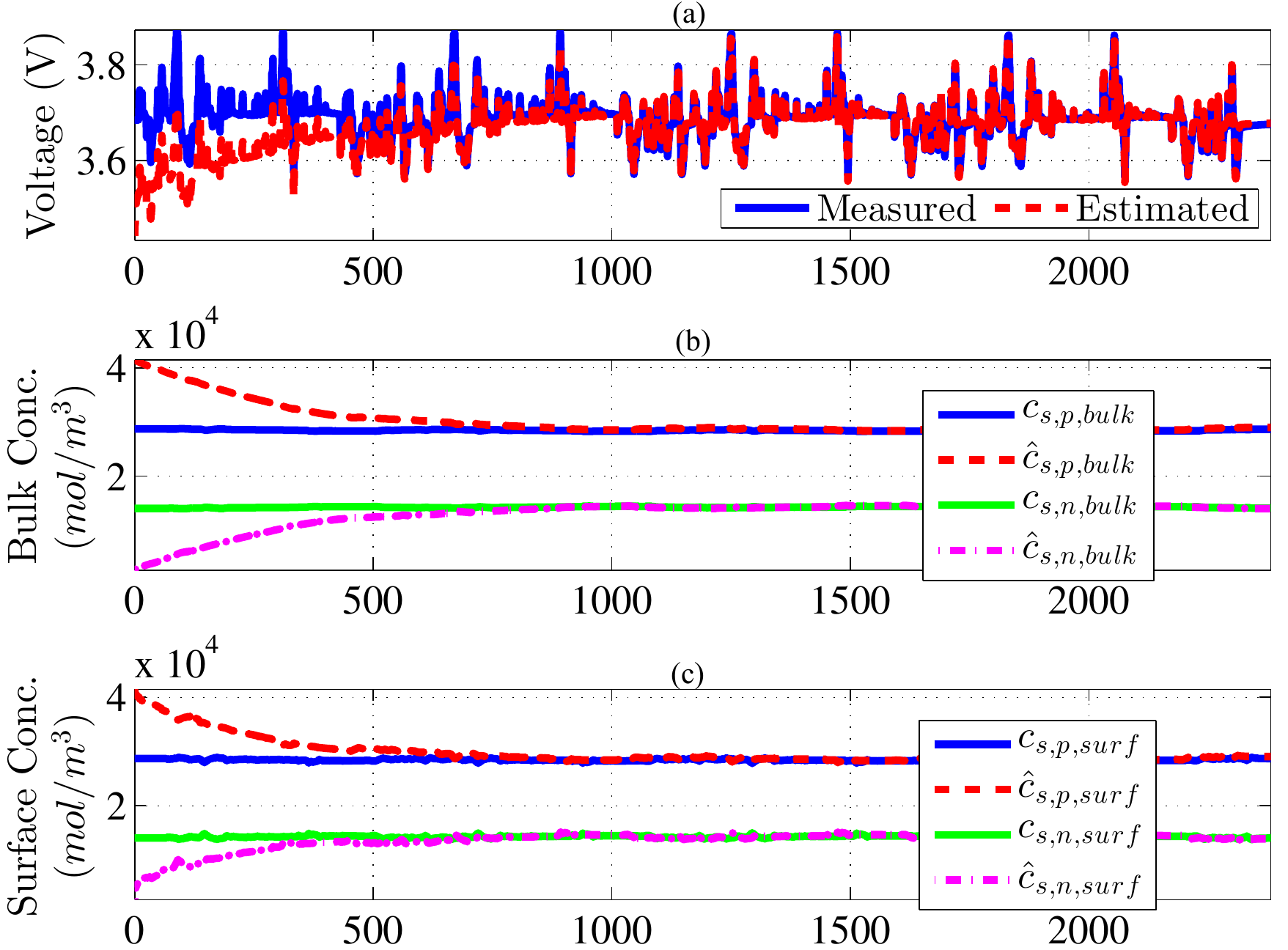}}
	\\
	{\includegraphics[width=8.5cm]{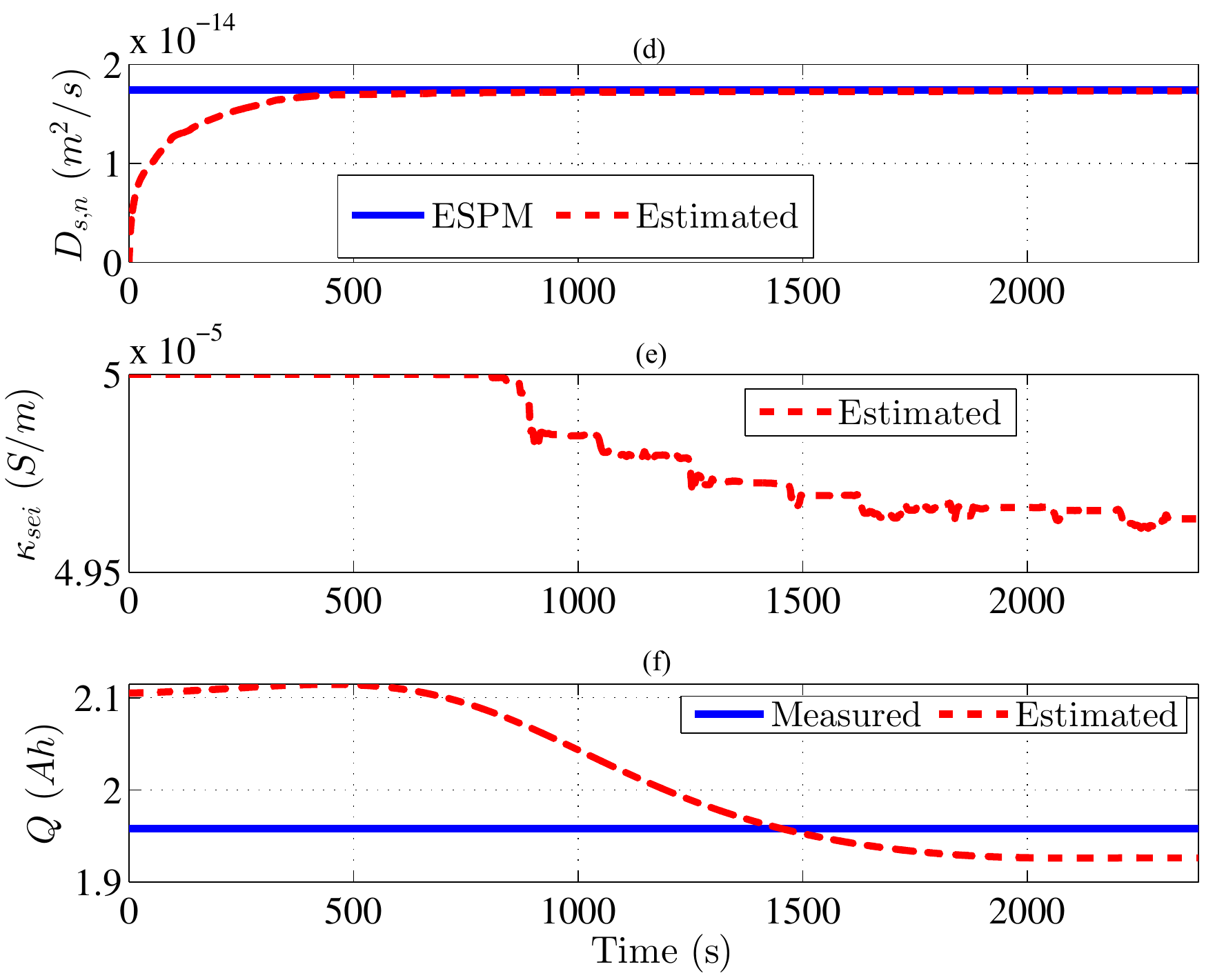}}
	\caption{ Performance evaluation of the interconnected adaptive observer for the US06 drive cycle of Cell $\#$ A. (a)Voltage estimation compared with experimental data, (b),(c),(d) Bulk, Surface concentration, and Diffusion coefficient estimation compared against ESPM values as the truth model, (e) Estimation of conductivity in the SEI layer, and (f) Capacity estimation validated against the measured capacity. Error in capacity estimate is less than 1$\%$.} 
	\label{fig:cell_A_result}
\end{figure}

\begin{figure}[tp]
	\centering
	{\includegraphics[width=8.5cm]{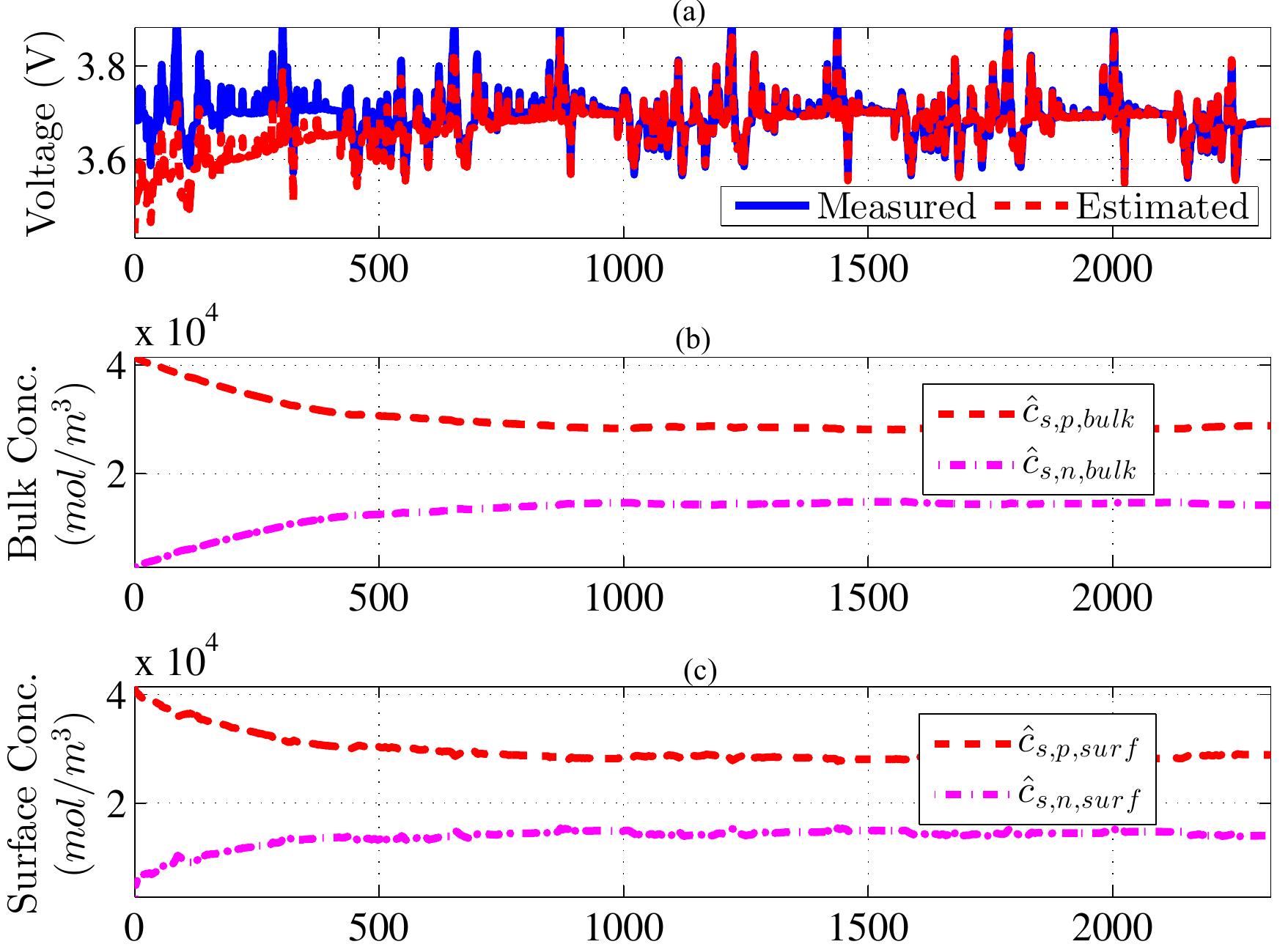}}
	\\
	{\includegraphics[width=8.5cm]{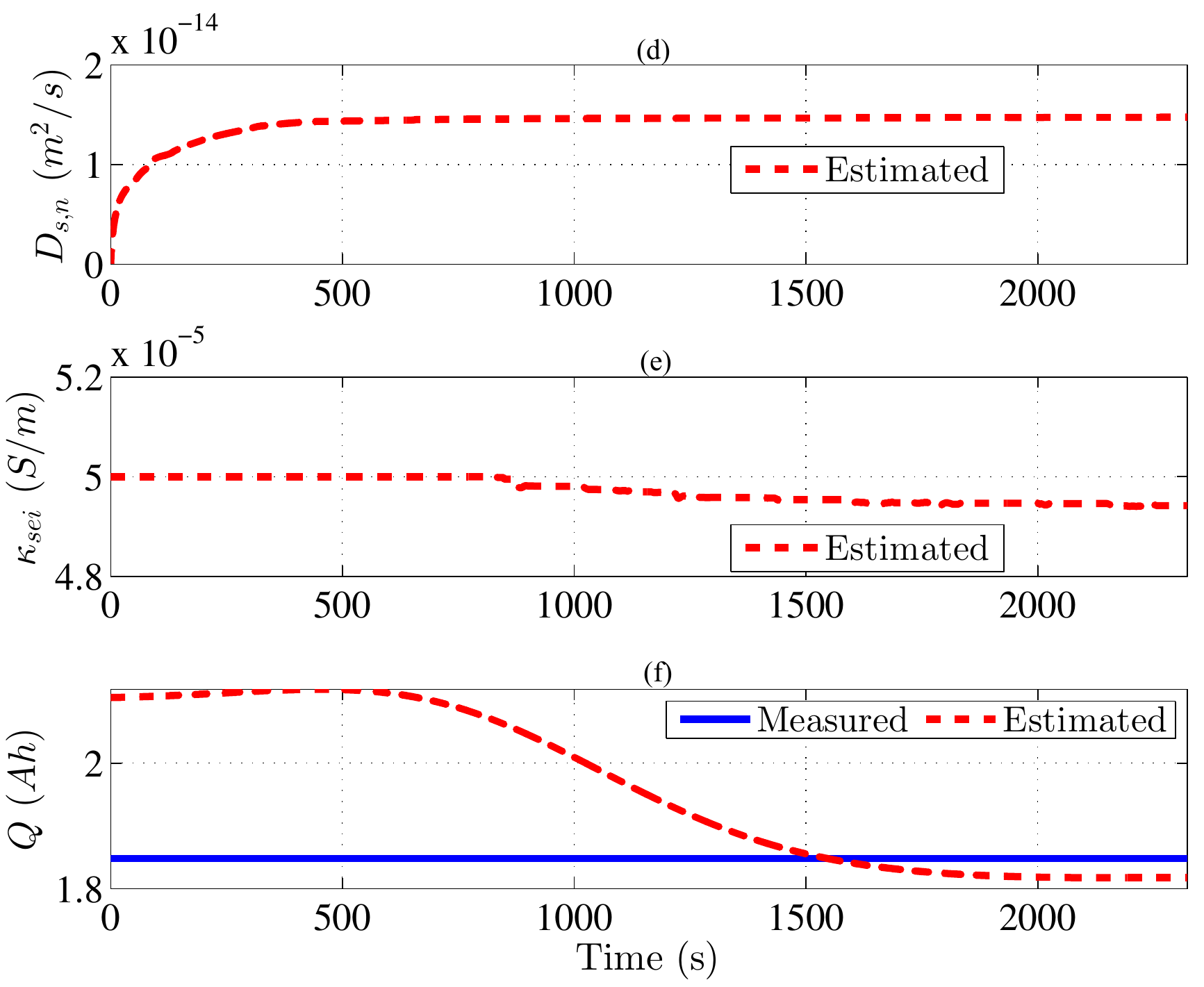}}
	\caption{Performance evaluation of the interconnected adaptive observer for the US06 drive cycle of Cell $\#$ B. (a)Voltage estimation compared with experimental data, (b),(c),(d), (e) Estimation of Bulk, Surface concentration, Diffusion coefficient, and Conductivity in the SEI layer, and (f) Capacity estimation validated against the measured capacity. Error in capacity estimate is less than 2$\%$.}
	\label{fig:cell_B_result}
\end{figure}

\begin{figure}[tp]
	\centering
	{\includegraphics[width=8.5cm]{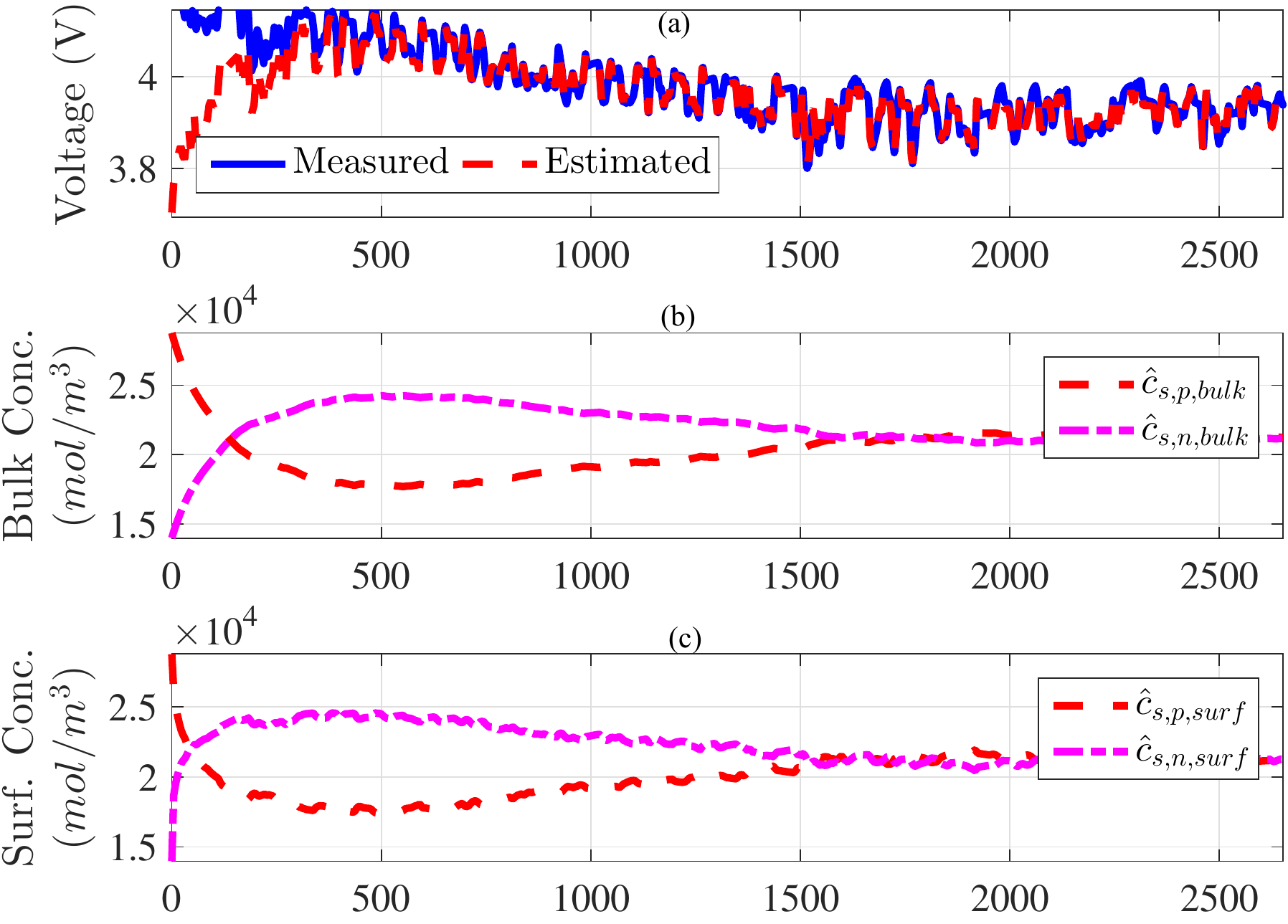}}
	\\
	{\includegraphics[width=8.5cm]{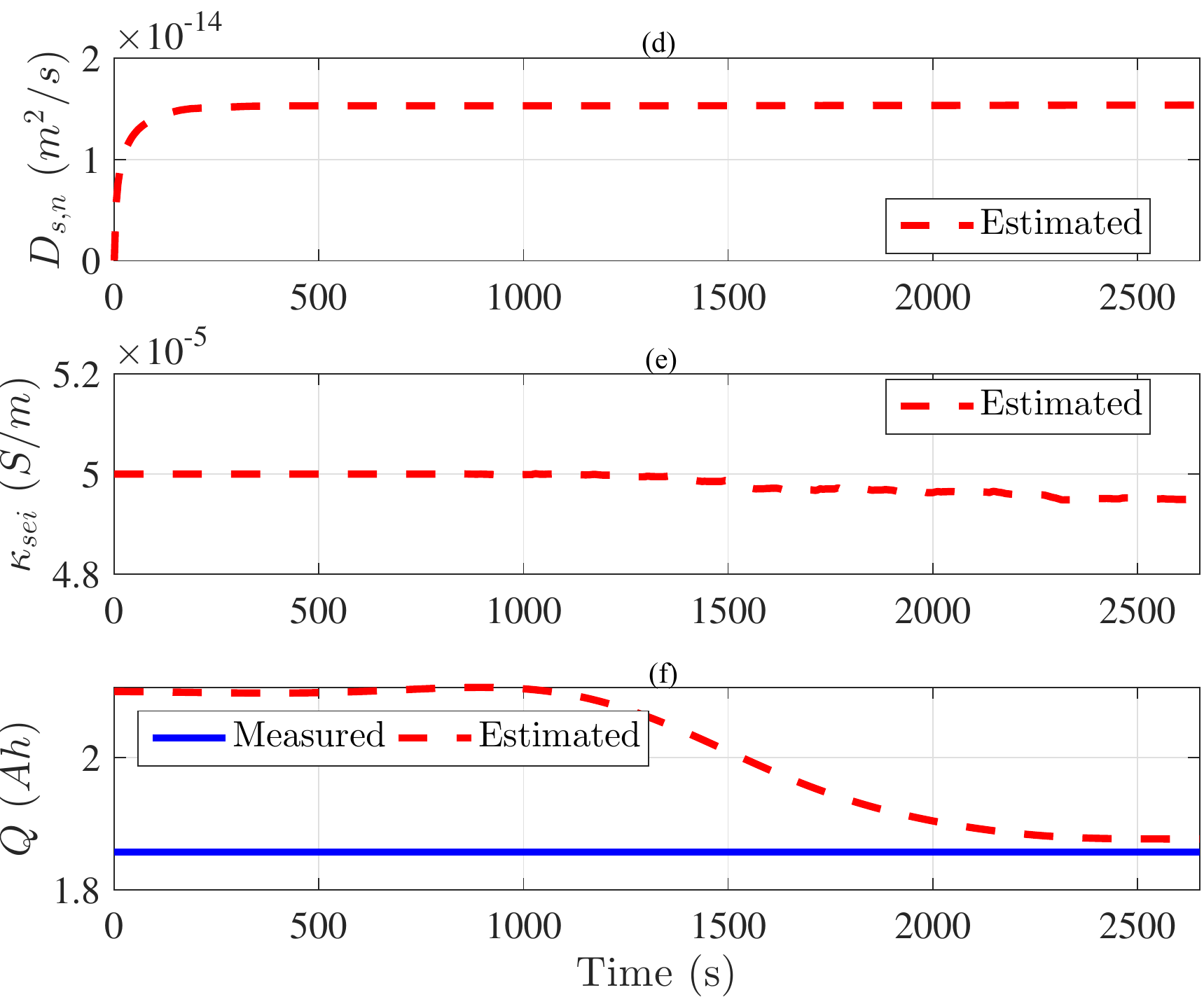}}
	\caption{Performance evaluation of the interconnected adaptive observer for the UDDS drive cycle of Cell $\#$ B. (a)Voltage estimation compared with experimental data, (b),(c),(d), (e) Estimation of Bulk, Surface concentration, Diffusion coefficient, and Conductivity in the SEI layer, and (f) Capacity estimation validated against the measured capacity. Error in capacity estimate is less than 2$\%$.}
	\label{fig:cell_B_result_UDDS}
\end{figure}

Cell $\#$ A is subjected to a US06 drive cycle derived from a Hybrid Electric Vehicle simulator and scaled for a single cell. The measured voltage and current data of Cell $\#$ A are fed as input to the interconnected adaptive observer. The lithium concentration states in both electrodes are initialized with an error of 45$\%$. The capacity of the observer is initialized to 2.1Ah, which is an error of 7.6$\%$ with respect to the true measured value of 1.95Ah. The diffusion coefficient is initialized as $\hat{D}_{s,n,ref} = 0.1\cdot D_{s,n,ref}$. The estimation performance is shown in Fig.~\ref{fig:cell_A_result}. The estimated capacity is 1.94Ah; which is well within 1$\%$ of its measured value. Since the actual value of SEI layer ionic conductivity is unknown, the convergence of the capacity estimate is taken as an indication of its convergence.
For the charge-sustaining US06 drive cycle, the measured voltage and current data of Cell $\#$ B are fed as input to the interconnected adaptive observer. The aged cell $\#$ B has lost approximately 6$\%$ of its capacity, as showin in Measured Capacity column in Table \ref{Cap_est}. The initialization error in states and parameters of the observer is same as the case of cell $\#$ A. The estimation performance is shown in Fig.~\ref{fig:cell_B_result}. Since the cell is aged, it is not possible to validate the non-measurable states and parameters like bulk, surface concentration and the anode diffusion coefficient against the ESPM in Section \ref{sec:Echem_model} which is for a fresh cell with nominal parameters. In this case, the estimation performance of voltage and capacity against experimentally measured values are taken as an indicator of the convergence of the internal states and parameters. The estimated capacity is 1.82Ah; which is within 2$\%$ of its measured value.
Next, a charge-depleting drive cycle (UDDSx2) derived for an electric vehicle and scaled to a single cell \cite{moura2014adaptive} is input to the aged cell $\#$ B. The initialization errors introduced in states and parameters of the observer are same as used in the previous case for the US06 profile. The estimation performance is shown in Fig.~\ref{fig:cell_B_result_UDDS}. The estimated cell capacity is again observed to be within 2$\%$ of its measured value.

\subsection{Estimation with measurement noise and sensor bias}
\textcolor{black}{
The measured current (US06 drive cycle) and voltage of Cell $\#$ A  is corrupted with a zero-mean Gaussian noise of 100mA and 25mV standard deviation, respectively. This is to mimic measurement noises introduced due to error in sensors or error in data transmission from the sensors. The adaptive observer is fed with the corrupted current and the corrupted voltage data to verify its robustness in capacity estimation. The estimation results as shown in Fig.~\ref{fig:cell_A_noise} is well within 2$\%$ of its measured value. Further, measured current and voltage data for Cell $\#$ A is corrupted by adding a constant bias of 10mA and 10mV to simulate a faulty un-calibrated sensor. The bias-induced current and voltage data from the experiment are supplied to the proposed adaptive observer. The capacity estimation, in this scenario, is also bounded within 2$\%$ of its real value, as shown in Fig.~\ref{fig:cell_A_bias}, indicating that the interconnected observer provides robust capacity estimates against sensor biases. %
\begin{figure}[tp]
	\centering
	{\includegraphics[width=8.5cm]{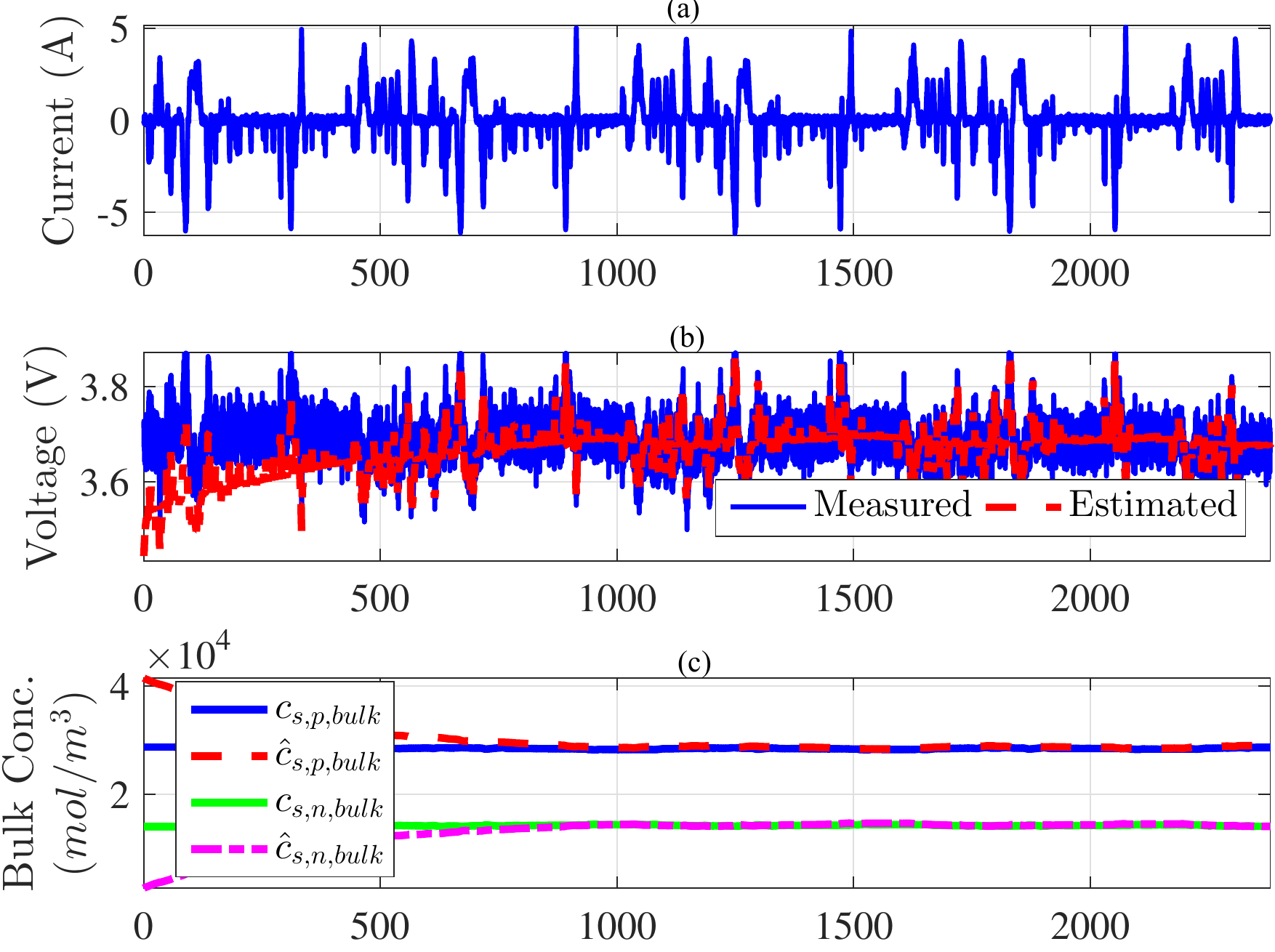}}
	{\includegraphics[width=8.5cm]{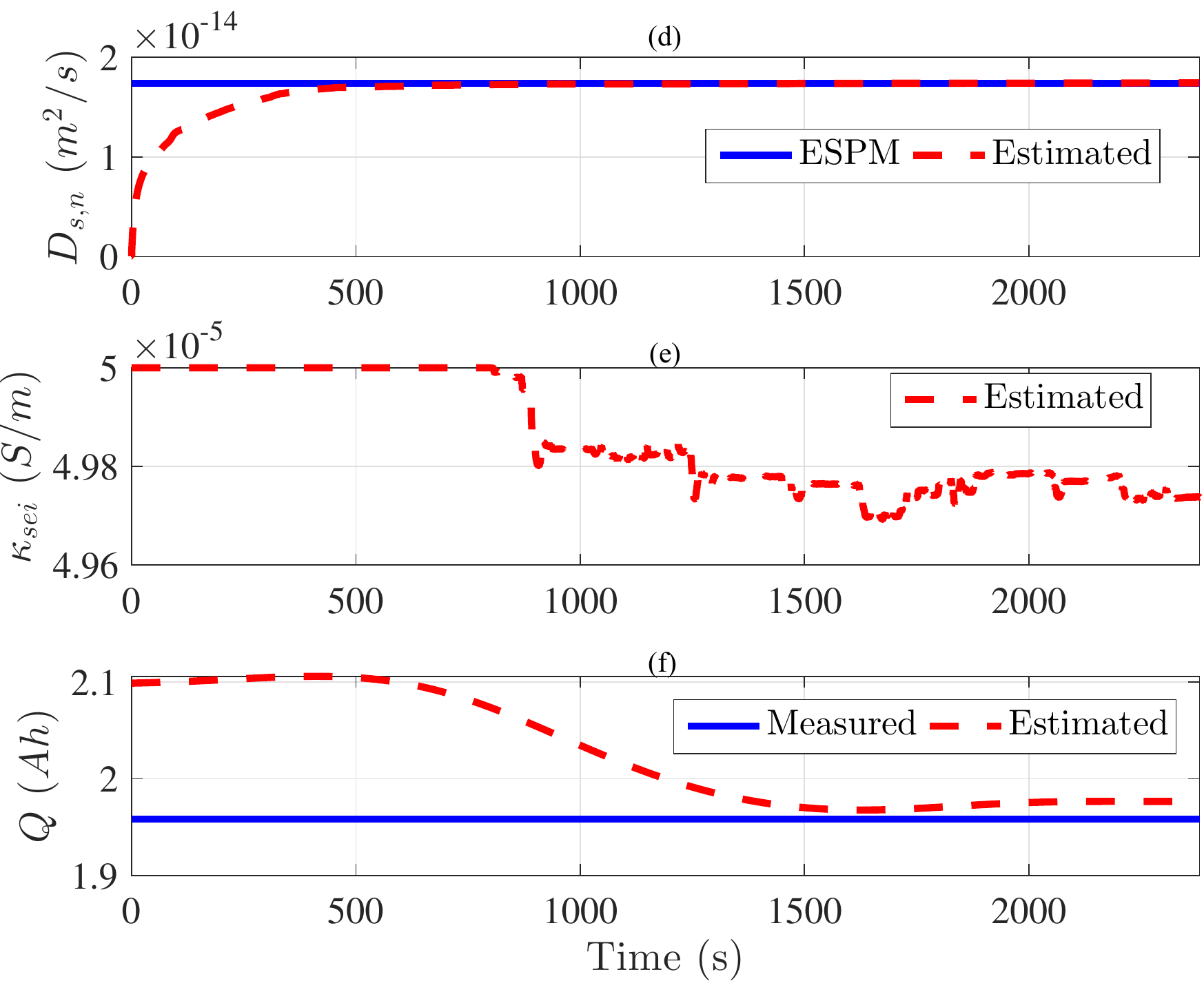}}
	\caption{Evaluation of the interconnected adaptive observer for the US06 drive cycle of Cell $\#$ A with the measured current and voltage corrupted with a zero-mean Gaussian noise of 100mA and 25mV standard deviation, respectively. (a) Corrupted US06 current input profile, (b) Voltage estimation compared with corrupted experimental data, (c),(d),(e) Estimation of Bulk concentration, Diffusion coefficient, and Conductivity in the SEI layer, and (f) Capacity estimation validated against the measured capacity. Error in capacity estimate is less than 2$\%$.}
	\label{fig:cell_A_noise}
\end{figure}
\begin{figure}[tp]
	\centering
	{\includegraphics[width=8.5cm]{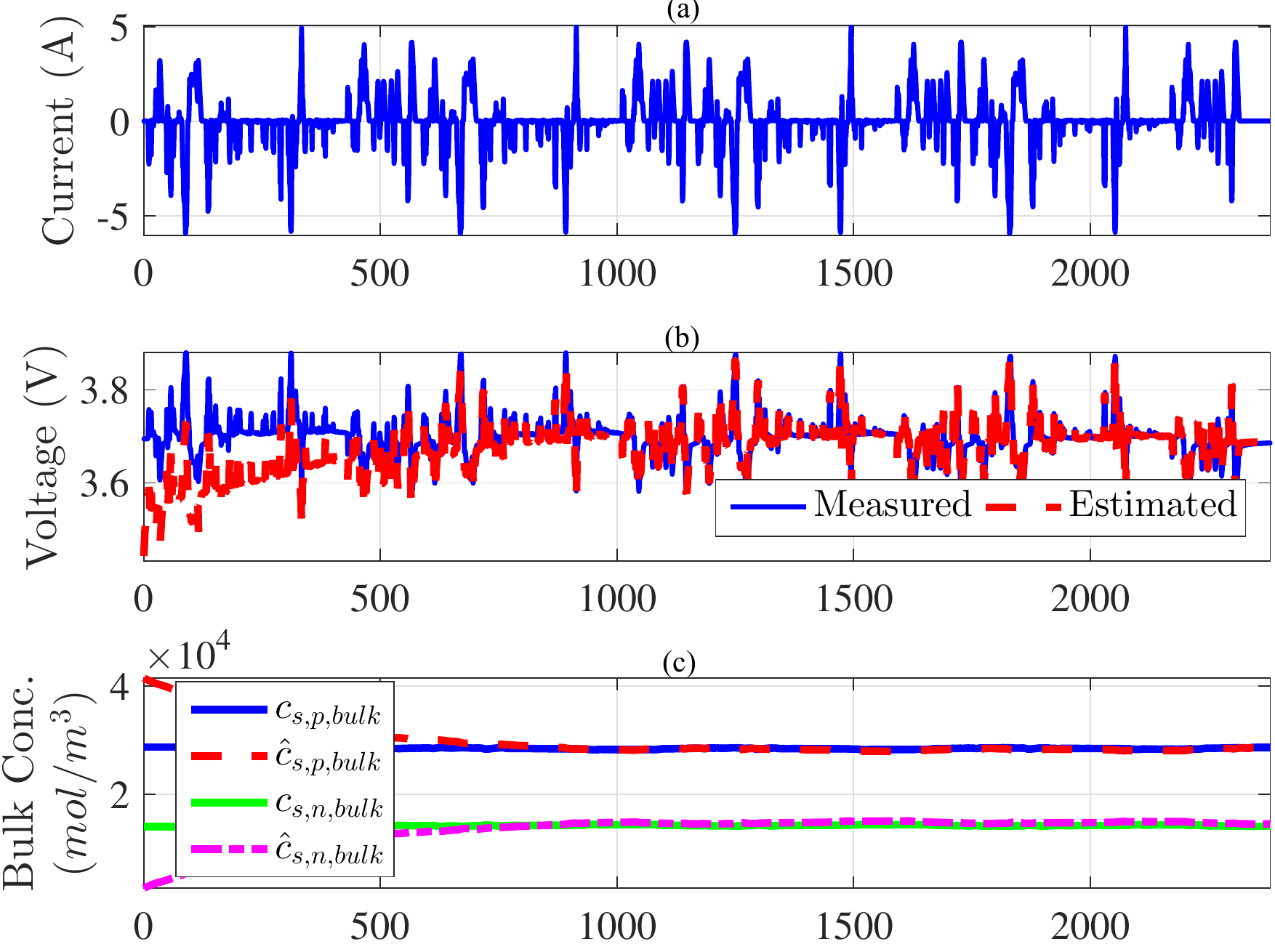}}
	{\includegraphics[width=8.5cm]{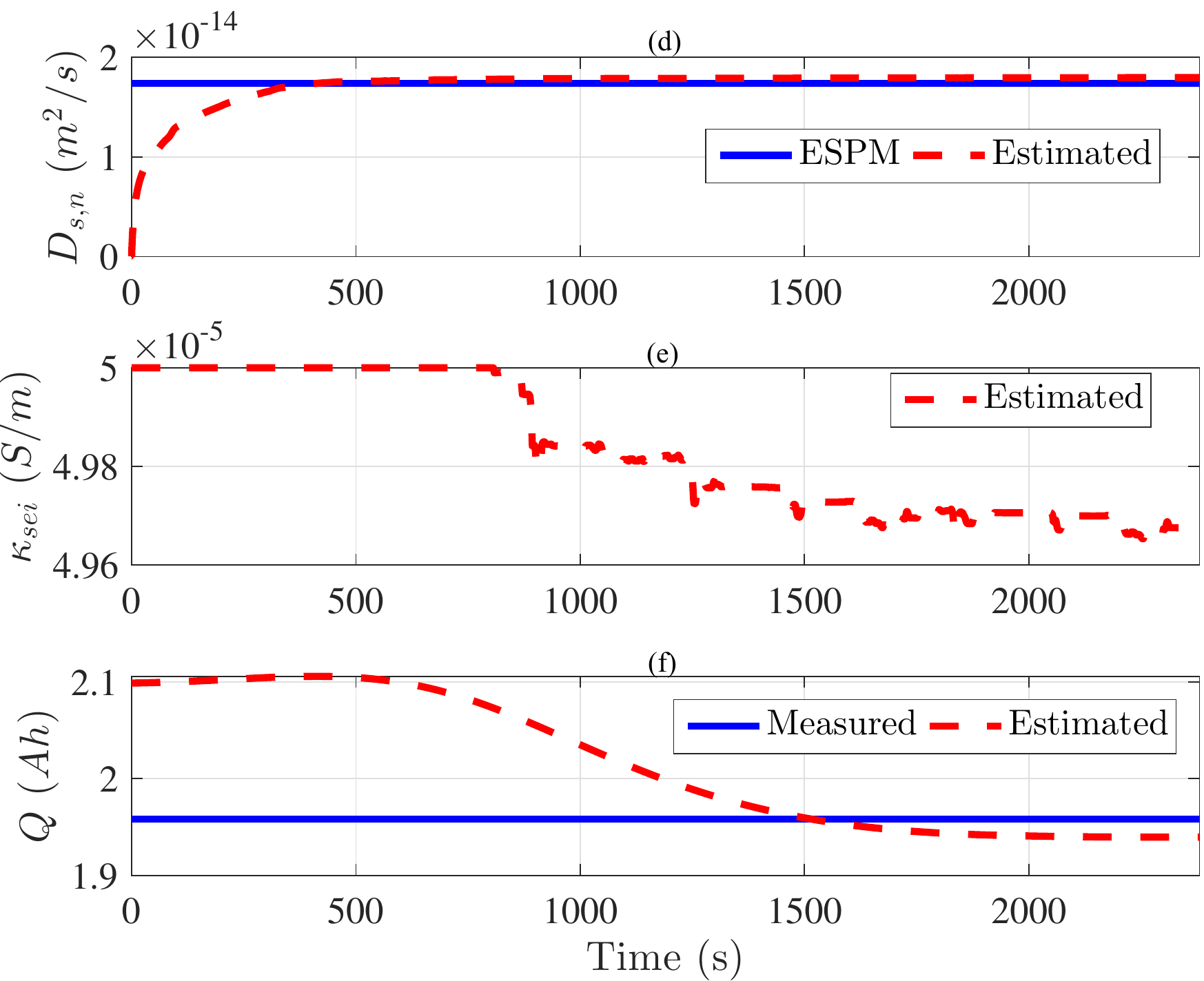}}
	\caption{Evaluation of the interconnected adaptive observer for the US06 drive cycle of Cell $\#$ A with the measured current and voltage corrupted with a constant bias 10mA and 10mV, respectively. (a) Corrupted US06 current input profile, (b) Voltage estimation compared with corrupted experimental data, (c),(d),(e) Estimation of Bulk concentration, Diffusion coefficient, and Conductivity in the SEI layer, and (f) Capacity estimation validated against the measured capacity. Error in capacity estimate is less than 2$\%$.}
	\label{fig:cell_A_bias}
\end{figure}
}
\section{Conclusion} \label{sec:Conclusion}
This paper addresses the issue of combined estimation of non-measurable critical battery variables such as lithium concentration and total cell capacity through an electrochemical model-based adaptive interconnected observer. Under the assumption that the SEI layer growth is the dominant aging mechanism, an adaptive interconnected observer is formulated by exploiting the dynamic relationship between capacity and power fade. A  model-based adaptive interconnected observer is proposed for combined estimation of lithium concentration in both electrodes, cell capacity, and aging-sensitive parameters such as anode diffusion coefficient and ionic conductivity in the SEI layer in real-time. Implementation results on different lithium-ion cells operating at varying stages of health show that the capacity estimates are bounded within 2$\%$ of their respective true value. Capacity estimates are found to be robust to measurement noise and sensor bias. 
\appendices
\section{Parameter Identification and Validation}\label{apd:Identification}
\begin{figure}[!ht]
\label{fig:cell_spec}{\includegraphics[width=\linewidth]{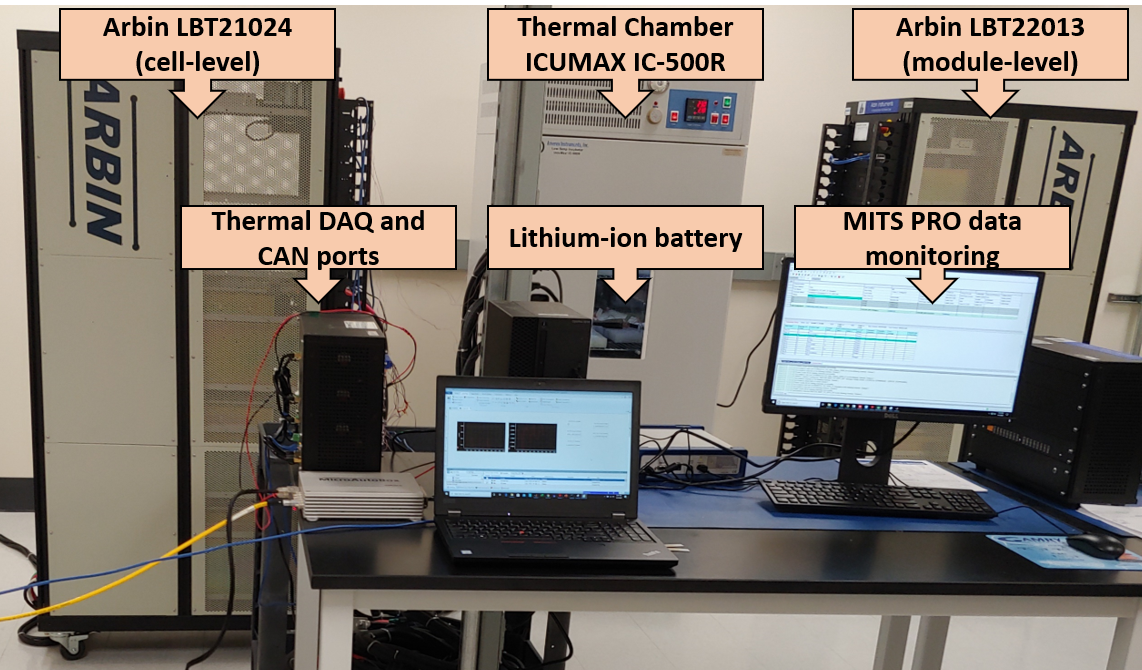}}%
	\hspace{0.5cm}
	\centering
	\label{fig:cell_spec2}{\includegraphics[width=3.5cm]{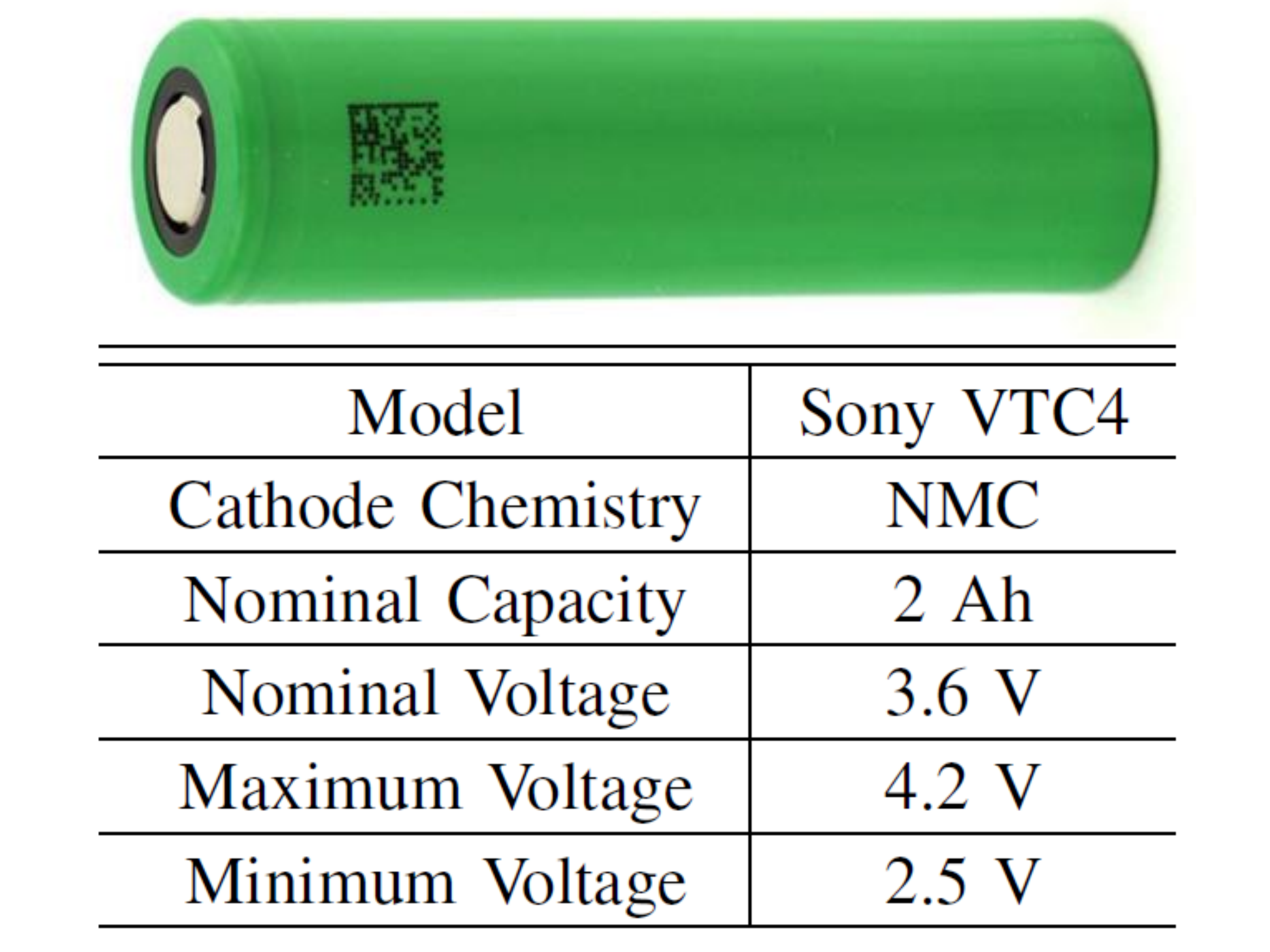}}%
	\caption{The experimental setup for battery testing and specifications of cylindrical 18650 2Ah NMC Lithium-ion cell used in the experiments.
	}
	\label{fig:setup_cell}
\end{figure}

The ESPM parameters in \eqref{eq:state_space} are identified from the experimental data collected over a cylindrical 2Ah NMC Lithium-ion cell. The experimental setup shown in Fig.~\ref{fig:setup_cell} includes two Arbin battery testing systems - capable of applying diverse current profiles to cells  - and a thermal chamber.  
%
The ESPM detailed in \eqref{eq:state_space} is characterized by a vector  $\lambda$ with 18 parameters to be identified
\begin{align}
	\lambda =& \big[c_{s,n,max}, c_{s,p,max}, D_{s,n}, D_{s,p}, R_{n}, R_{p},\nonumber \\ \nonumber
	& A, L_{n}, L_{p}, \epsilon_{n}, \epsilon_{p}, k_{n}, k_{p}, R_{l}, L_{s}, \epsilon_{e,s}, \epsilon_{n,f}, \epsilon_{p,f}   \big]^{T}. \nonumber
\end{align}
The identification of the parameter vector $\lambda$ is achieved by fitting the ESPM output voltage to the measured voltage data. However, it is well understood that electrochemical models, such as ESPM, are nonlinear in parameters and all the parameters may not be uniquely identifiable from the output voltage \cite{forman2012genetic}. The need for identifying 18 parameters leads to over-parameterization, especially when a small subset of parameters are sufficient to predict the behavior of the model. Thus, the conventional parameter identification technique involving only a solitary objective function of minimizing error between simulated output and measured output voltage reveals parameter identifiability issues.  In this work, an attempt is made to enhance the existing identification technique by incorporating virtual measurements into the objective function. The virtual measurement is in the form of SOC computed using Coulomb Counting method from the measured current data. For an off-line parameter identification study conducted under perfectly controlled laboratory conditions, it is safe to assume that the initial state of charge and temperature are known, and that  the current measured by the Arbin is highly accurate. The identifiability of ESPM parameters is maximized by solving a multi-objective optimization problem that minimizes the combination of following objectives: 1)  $J_{1}:$ error between measured and simulated voltage, 2) $J_{2}:$ error between Coulomb Counting SOC and cathode bulk SOC computed from volume averaging of cathode concentration,  and 3) $J_{3}:$ error between Coulomb Counting SOC and anode bulk SOC computed from volume averaging of anode concentration\footnote{Note that the bulk SOC of both electrodes is assumed to be same because the cell is fresh and the assumption of conservation of lithium moles between both electrodes is valid.}. The advantages of the aforementioned multi-objective optimization is verified by analyzing the identifiability of the ESPM parameters with respect to the measured output voltage and virtually measured bulk SOC of both electrodes. In this work, the identifiability analysis is performed in two steps: (a) Local sensitivity analysis, and (b) Correlation analysis. 

\subsection{Sensitivity Analysis} 
\begin{figure*}[!t]\centering
	\includegraphics[width=16cm]{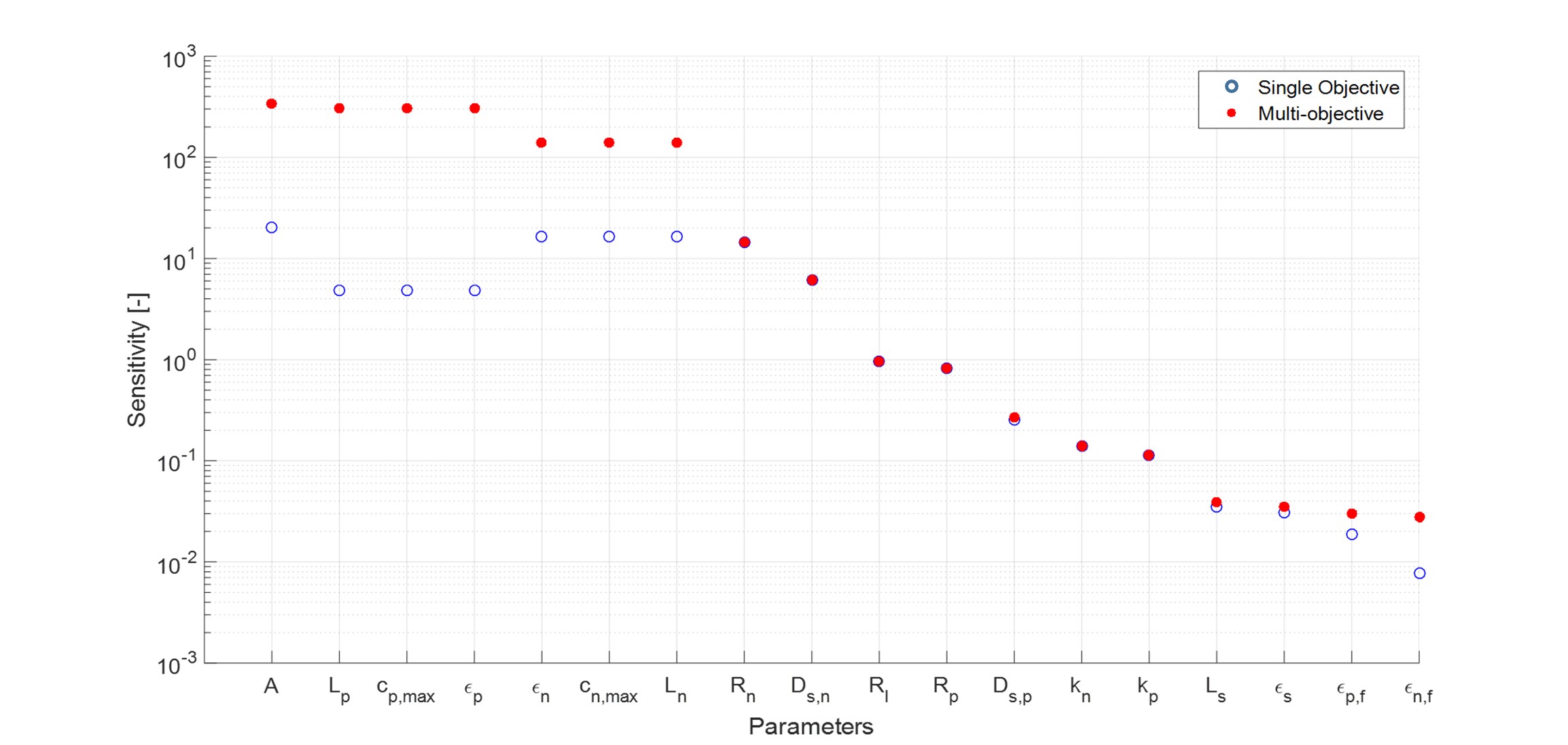}
	\caption{Ranked sensitivity comparison for EPSM parameters with single and multi-objective optimization functions. }\label{fig:sensitivity_plot}
\end{figure*}

The response of the ESPM outputs (voltage and bulk SOC) to changes in each parameter in $\lambda$ quantifies the sensitivity of the model output to the specific parameter. The nominal values for the parameters in $\lambda$ are taken from \cite{tanim2015temperature}. The sensitivity is computed as
{\scriptsize \begin{align}
	\label{eq:sensitivity_matrices}
	S &=  \begin{bmatrix} \dfrac{\partial y_{1}}{\partial \lambda_{1}}\left(t_{1}\right) & \hdots & \dfrac{\partial y_{1}}{\partial \lambda_{j}}\left(t_{1}\right) \\%
		\vdots &  \dots & \vdots\\%
		\dfrac{\partial y_{1}}{\partial \lambda_{1}}\left(t_{k}\right) & \hdots & \dfrac{\partial y_{1}}{\partial \lambda_{j}}\left(t_{k}\right) \\%
		\dfrac{\partial y_{2}}{\partial \lambda_{1}}\left(t_{1}\right) & \hdots & \dfrac{\partial y_{2}}{\partial \lambda_{j}}\left(t_{1}\right) \\%
		\vdots &  \dots & \vdots\\%
		\dfrac{\partial y_{m}}{\partial \lambda_{1}}\left(t_{k}\right) & \hdots & \dfrac{\partial y_{m}}{\partial \lambda_{j}}\left(t_{k}\right) \\%
	\end{bmatrix},  
\end{align}}
where $m$ is the number of outputs (cell voltage and bulk SOC of both electrodes, $m = 3$), $k$ is the number of total samples available, $j$ is the number of parameters, and $S \in \mathbb{R}^{\left(k \times m\right) \times j}$. %
In order to compare the sensitivities of all parameters, the Euclidean norm of every column of the sensitivity matrix corresponding to each parameter $(||S_{:,j}||)$ is computed. The parameters sorted as per their sensitivities $(||S_{:,j}||)$ with respect to multi-outputs (voltage and bulk SOC) and single-output (voltage) is compared on a log scale in Fig~\ref{fig:sensitivity_plot}. From Fig.~\ref{fig:sensitivity_plot}, it is verified that incorporating another output in the form of SOC, indeed, improves the sensitivity of the parameters. 

\subsection{Correlation Analysis} 
\begin{figure*}[!ht]\centering
	\centering
	{\includegraphics[width=9cm]{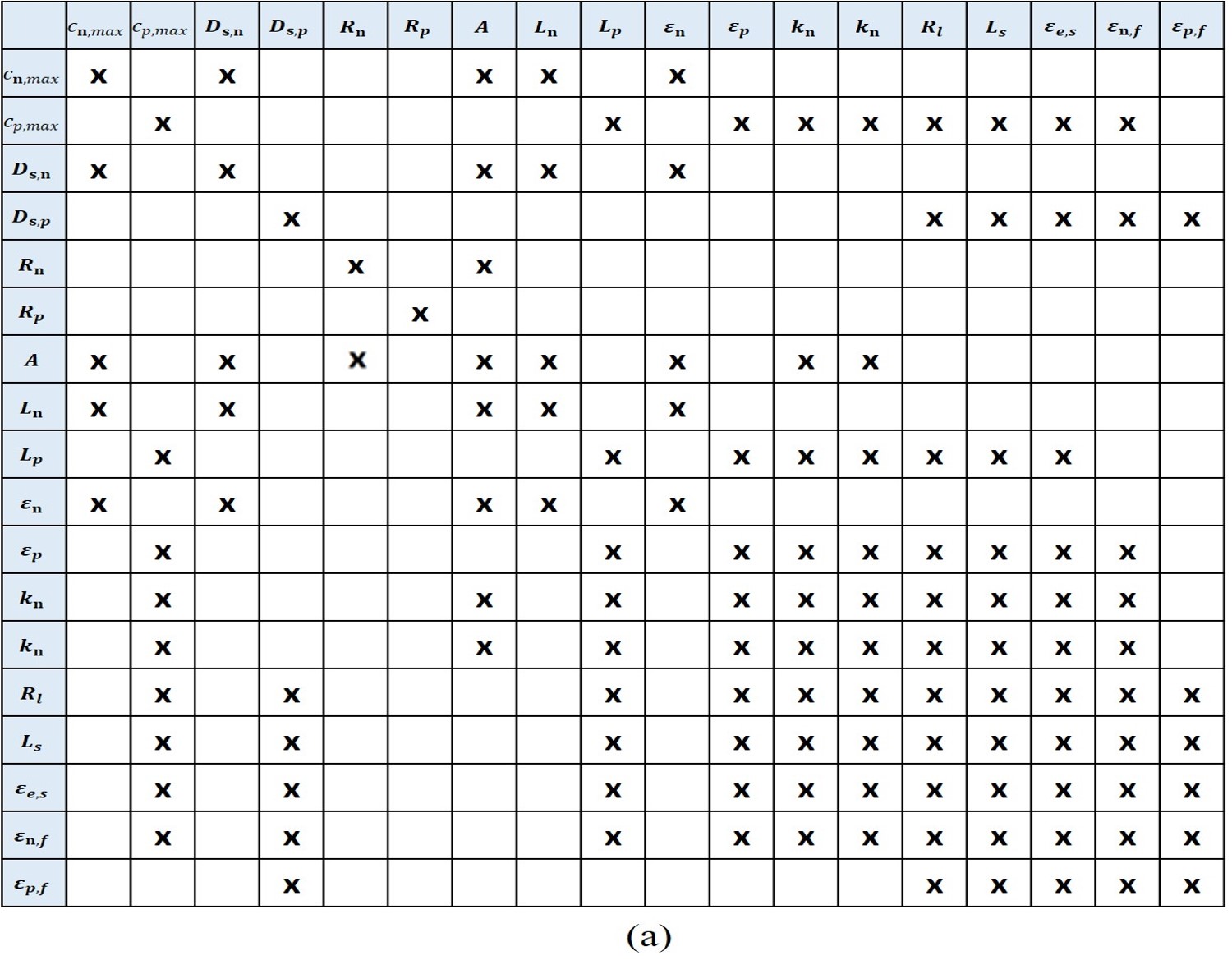}}
	{\includegraphics[width=9cm]{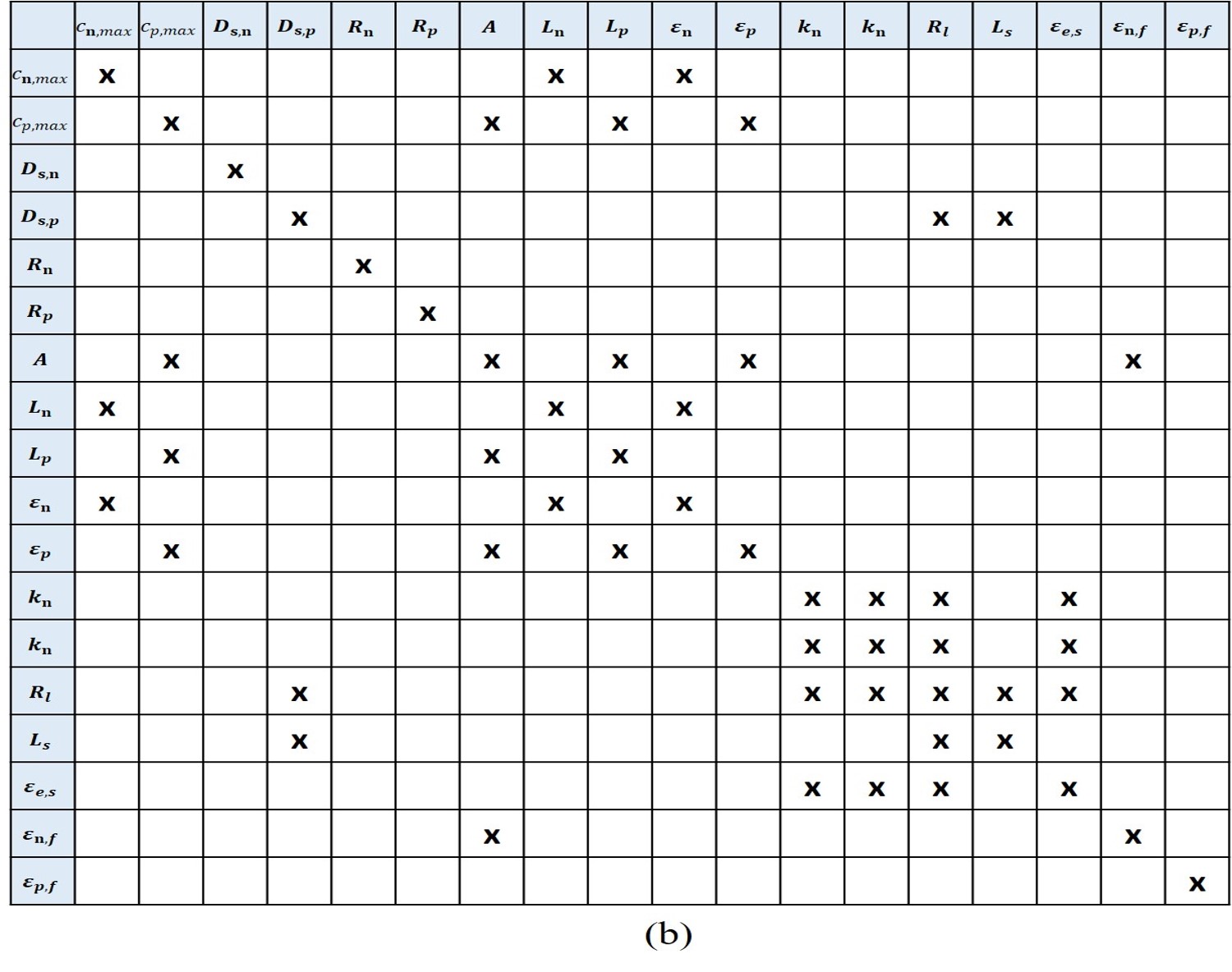}}
	\caption{Correlation analysis for ESPM parameters (a) for single objective function, and (b) for multi-objective function. The times symbol ($\times$)  represents values of $\bar{C}_{i,j} > 0.8$ and hence indicates that parameters are correlated.}
\label{fig:correlation_analysis}
\end{figure*}

\begin{figure}[!t]\centering
	\includegraphics[width=\linewidth]{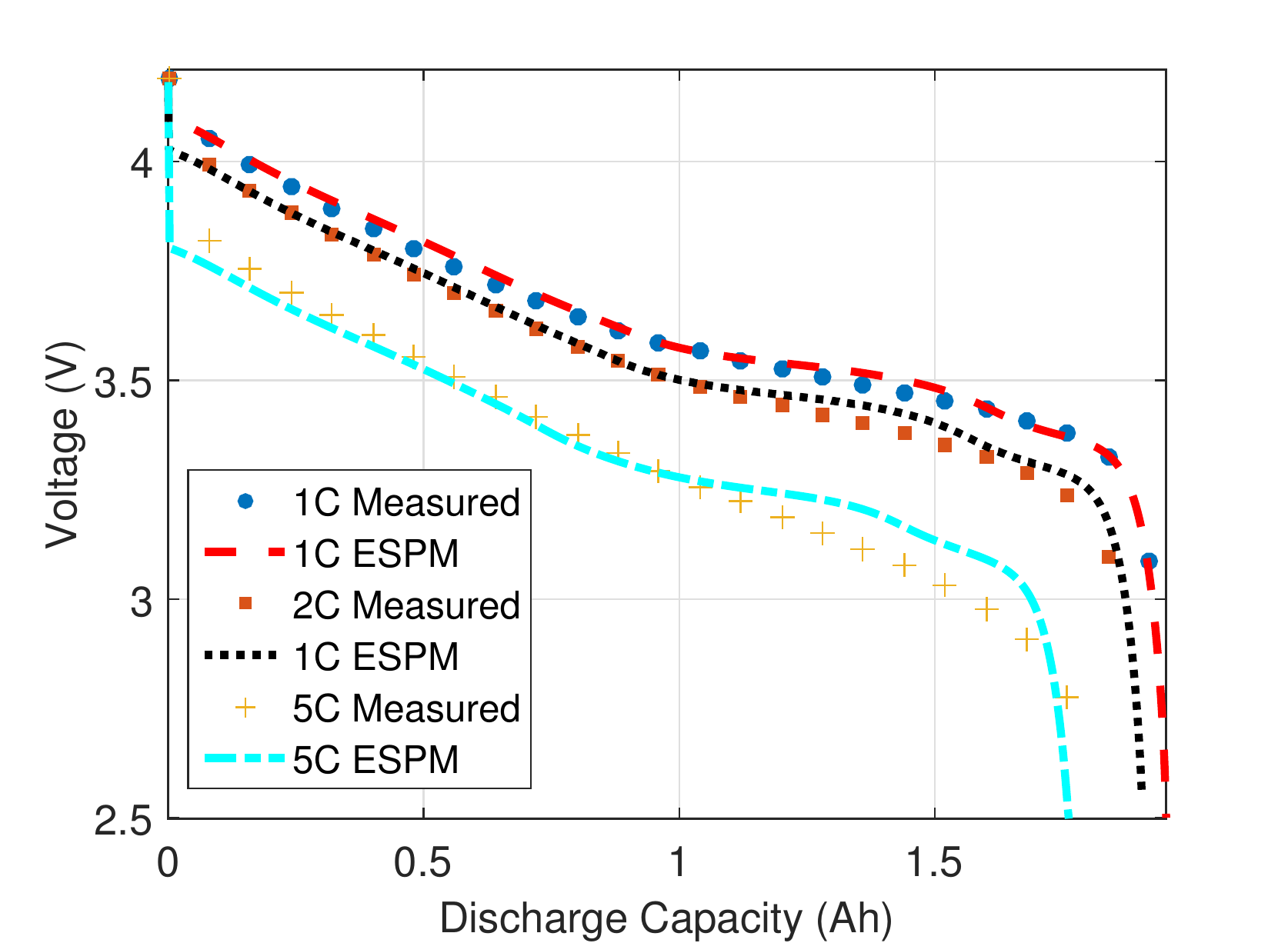}
	\caption{Comparison of ESPM output voltage with measured cell voltage for 1C, 2C, and 5C constant current discharge cycles at $23^{o}C$.} \label{fig:espm_const_current}
\end{figure}

Despite the improved sensitivity of some parameters (see  Fig.~\ref{fig:sensitivity_plot}), it is important to verify if they can be uniquely identified from the available outputs. Correlation analysis is performed where the linear dependence of the sensitivity matrix columns is computed as
{\scriptsize \begin{align}
	\label{eq:correlation_matrices}
	\bar{C} &=  \begin{bmatrix} \bar{C}_{1,1} & \bar{C}_{1,2} & \hdots & \bar{C}_{1,j} \\%
		\bar{C}_{2,1} & \bar{C}_{2,2} & \hdots & \bar{C}_{2,j} \\%
		\vdots &  \vdots & \hdots & \vdots \\%
		\bar{C}_{j,1} & \bar{C}_{j,2} & \hdots & \bar{C}_{j,j} \\%
	\end{bmatrix},  
\end{align}}
where each element in the correlation matrix $\bar{C}$ is computed as
\begin{equation}
	\bar{C}_{i,j} =  \dfrac{\langle S_{:,i}, S_{:,j} \rangle}{||S_{:,i}|| ||S_{:,j}||}.
\end{equation}
Essentially, if changes in different parameters result in the same response in the outputs, their respective sensitivity columns will be similar or linearly dependent. Hence, values of $\bar{C}_{i,j}$ close to 1 or -1 indicate linear dependency between parameters and hence they cannot be identified uniquely from the outputs. In this work, the threshold value for $\bar{C}_{i,j}$ to indicate correlation is taken to be $0.8$. The correlation analysis for parameter identification with a solitary objective function is shown in Fig.~\ref{fig:correlation_analysis}.a and with multi-objective function is shown in Fig.~\ref{fig:correlation_analysis}.b. 
The results from identifiability analysis shows that although including SOC as a virtual measurement improves sensitivity and reduces correlation between parameters, there is still not a single parameter that is uniquely identifiable. Hence, a subset of parameters is selected based on the ranked sensitivity list and the correlation analysis table that can sufficiently characterize the ESPM without leading to over-parametrization. The vector consisting of parameters that can be uniquely identified from the outputs is denoted by $\lambda^{*}$. Each parameter with a sensitivity value higher than a threshold value of $||S_{:,j}||>0.2$ is considered for the parameter subset selection procedure. Firstly, the most sensitive parameter, the cell cross-sectional area $A$, is automatically selected in the subset parameter vector $\lambda^{*}$. Next, the second ranked sensitive parameter is checked for correlation with $A$. If it is correlated, then the parameter is fixed at its nominal value, taken from the literature \cite{tanim2015temperature}. If the parameter is not correlated to $A$, then it enters the subset parameter vector $\lambda^{*}$ as a parameter that can be uniquely identified. The process continues until every parameter is checked. Based on the subset selection procedure, the set of parameters that can be uniquely identified is given as $\lambda^{*} = \big[A, \epsilon_{n}, R_{n}, D_{s,n}, R_{l}, R_{p}, D_{s,p} \big]^{T}$. 
Note that this analysis is specific to the input current profile used. 
The identifiability analysis procedure reduces the number of parameters that need to be identified from 18 to 7, thereby reducing over-parameterization. 
The multi-objective optimization problem  is then formulated as follows:
\begin{equation*}
	\begin{aligned}
		& \underset{\lambda^{*}_{min} < \lambda^{*} < \lambda^{*}_{max}}{\text{argmin}}
		& & J_{1} + J_{2} + J_{3} \\
		& \text{subject to:}
		& & \dot{x}_{1,k} = A_{11} \left( \lambda^{*} \right) x_{1,k} + B_{1}  \left( \lambda^{*} \right)u_{k}, \\
		&&& \dot{x}_{2,k} = \lambda^{*}_{4}\bar{A}_{22} \left( \lambda^{*} \right) x_{2,k} + B_{2} \left( \lambda^{*} \right)u_{k}, \\
		&&& \dot{x}_{3} = 0, \\
		&&& \dot{x}_{4,k} = f_{e} \left(x_{4},u,\lambda^{*} \right) \nonumber \\
		&&& y_{k}= h_{1}(x_{1,N,k},u) - h_{2}(x_{2,N,k},u) - \\
		&&& \qquad h_{4}(x_{4,k},u) - \lambda^{*}_{5}u_{k} -\nonumber \\ \nonumber
		&&& \qquad h_{3}(x_{3,k})u_{k} + \left(x_{3,k} - Q_{0}\right)\theta_{2}u_{k} \\
		&&& Q_{0}=\dfrac{F   L_{p} \epsilon_{p} c_{s,p,max} \left(\theta_{p,100\%}- \theta_{p,0\%} \right)  \lambda^{*}_1}{3600}, \\
	\end{aligned}
\end{equation*}
where $\lambda^{*}$ is the vector  containing the parameters to be identified, and $u_{k}$ is the  experimentally measured input.
 Recall that the identification procedure is carried out for a fresh cell, hence the terms due to aging $h_{3}(x_{3,k})u_{k} + \left(x_{3,k} - Q_{0}\right)\theta_{2}u_{k}$ are $0$ because $x_{3} = Q_{0}$\footnote{Remaining parameters in $\lambda$ that are not being identified assume nominal values from the literature \cite{tanim2015temperature}.}. The multi-objective constrained optimization problem is solved using Genetic Algorithm over the experimentally collected voltage and current data. %
The RMS error in voltage prediction by ESPM (Fig. \ref{fig:espm_const_current}) is as follows: RMS=17mV at 1C (identification), RMS=30.4mV at 2C and RMS=69.1mV at 5C (validation). 

%
%
%
%
%
%


\ifCLASSOPTIONcaptionsoff
  \newpage
\fi



\bibliographystyle{IEEEtran}
\bibliography{autosam}
%
%
%

%
\begin{IEEEbiography}[{\includegraphics[width=1in,height=1.25in,clip,keepaspectratio]{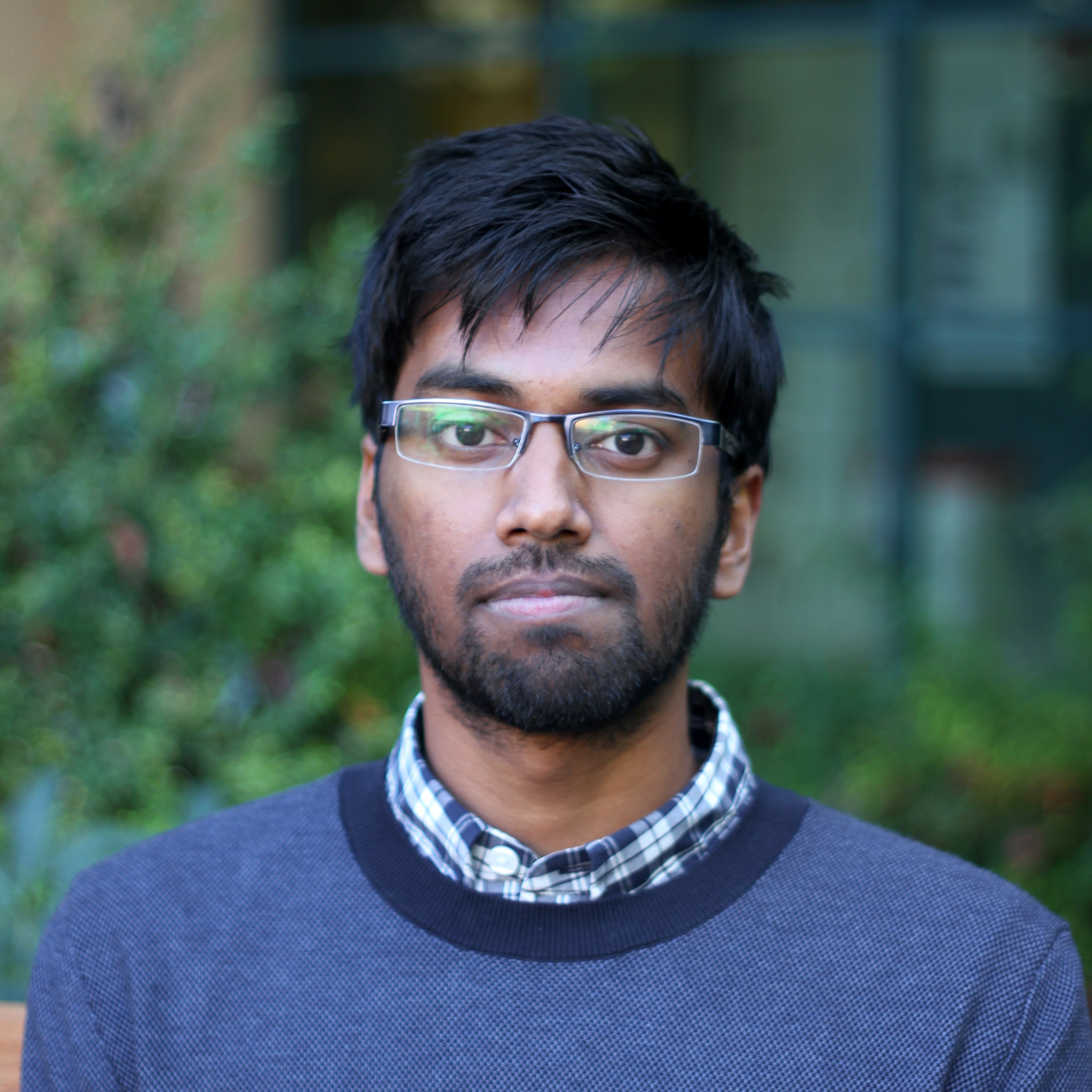}}]
	{Anirudh Allam} (Graduate Student Member, IEEE) received the B.E. degree in electronics and telecommunication
	engineering from the University of Pune,
	Pune, India, in 2010, and the M.S. degree in automotive engineering from Clemson University,	Clemson, SC, USA, in 2015. He is currently pursuing	the Ph.D. degree with the Department of Energy Resources Engineering, Stanford University,	Stanford, CA, USA.
	
	His research interests include estimation, control, and degradation modeling of electrochemical energy storage systems.
\end{IEEEbiography}

\begin{IEEEbiography}[{\includegraphics[width=1in,height=1.25in,clip,keepaspectratio]{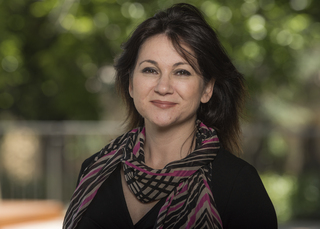}}]
	{Simona Onori} (Senior Member, IEEE) received the Laurea degree in computer science and engineering	from the University of Rome ``Tor Vergata'' Rome,
	Italy, in 2003, the M.S. degree in electronics and	communications engineering from The University of
	New Mexico, Albuquerque, NM, USA, in 2005, and the Ph.D. degree in control engineering from the University of Rome ``Tor Vergata'' in 2007.
	
	She is currently an Assistant Professor with the Energy Resources Engineering Department, Stanford University, Stanford, CA, USA. Her research
	focuses on modeling and control in sustainable transportation, clean energy,	and secondary life battery areas.
	
	Dr. Onori was a recipient of the 2019 Board of Trustees Award for Excellence, Clemson University, the 2018 Global Innovation Contest Award from LG Chem, the 2018 SAE Ralph R. Teetor Educational Award, and the 2017 NSF CAREER Award. She has been serving as the Editor-in-Chief for	the SAE International Journal of Electrified Vehicles since 2020. She is a Distinguished Lecturer of the IEEE Vehicular Technology Society.
\end{IEEEbiography}

%
%
%




\end{document}